\documentclass[
reprint,
superscriptaddress,
amsmath,amssymb,
aps, prx,
]{revtex4-2}

\usepackage[utf8]{inputenc}
\usepackage[T1]{fontenc}
\usepackage{graphicx} 
\usepackage{amsthm,amsfonts}
\usepackage{physics}
\usepackage{booktabs}
\usepackage[dvipsnames]{xcolor}
\usepackage{hyperref}
\usepackage{soul}
\usepackage{siunitx}

\newcommand{\E}{\mathrm{e}}
\newcommand{\I}{\mathrm{i}}

\begin{document}

\title{Expanding the neutral atom gate set: Native iSWAP and exchange gates from dipolar Rydberg interactions}

\author{Pedro Ildefonso}
\thanks{Contact author: p.ildefonso@parityqc.com}
\affiliation{Institute for Theoretical Physics, University of Innsbruck, A-6020 Innsbruck, Austria}
\affiliation{Parity Quantum Computing GmbH, Rennweg 1, Top 314, A-6020 Innsbruck, Austria}

\author{Andrew Byun}
\affiliation{Institute for Theoretical Physics, University of Innsbruck, A-6020 Innsbruck, Austria}

\author{Aleksei Konovalov}
\affiliation{Institute for Theoretical Physics, University of Innsbruck, A-6020 Innsbruck, Austria}

\author{Javad Kazemi}
\affiliation{Parity Quantum Computing Germany GmbH, Schauenburgerstraße 6, 20095 Hamburg, Germany}

\author{Michael Schuler}
\affiliation{Parity Quantum Computing GmbH, Rennweg 1, Top 314, A-6020 Innsbruck, Austria}

\author{Wolfgang Lechner}
\affiliation{Institute for Theoretical Physics, University of Innsbruck, A-6020 Innsbruck, Austria}
\affiliation{Parity Quantum Computing GmbH, Rennweg 1, Top 314, A-6020 Innsbruck, Austria}
\affiliation{Parity Quantum Computing Germany GmbH, Schauenburgerstraße 6, 20095 Hamburg, Germany}
\affiliation{Parity Quantum Computing France SAS, 10 Avenue de Kle\'eber, 75016 Paris, France}

\date{\today}

\begin{abstract}
\noindent We present a native realization of iSWAP and parametrized \textit{exchange} gates for neutral-atom quantum processing units. Our approach leverages strong dipole-dipole interactions between two different dipole-coupled Rydberg states, employing optimal control techniques to design high-fidelity, time-efficient gate pulses. 
To minimize experimental complexity, we utilize global driving fields acting identically on all atoms and apply pulse smoothing techniques. 
While detrimental van der Waals interactions pose a significant challenge, we demonstrate that for both $^{133}$Cs, as a representative alkali atom, and $^{88}$Sr, an alkaline-earth species, high-fidelity pulses can nevertheless be obtained over a broad range of parameters. 
We identify candidate protocols with reduced susceptibility to noise and analyze their performance under realistic conditions, accounting for atomic motion, Rydberg decay, and experimentally motivated laser frequency and intensity noise. 
Crucially, we demonstrate that in both Alkali and alkaline-earth-based systems, we can obtain fast iSWAP gates with fidelities of $99.9\%$ under realistic experimental conditions.
These results pave the way for expanding the neutral-atom gate set beyond conventional Rydberg-blockade-based entangling gates.
\end{abstract}

\maketitle

\section{Introduction}
Neutral-atom quantum processing units (QPUs) have become a leading platform for quantum computing due to their unique capabilities, including high qubit numbers, flexible connectivity, and long coherence times. 
Their scalability has been demonstrated by capturing and coherent manipulation of up to thousands of individual atoms~\cite{Gyger2024, Norcia2024, Manetsch2024, Chiu2025}.
High-fidelity quantum gates have been demonstrated for several atomic species and qubit encodings~\cite{Bluvstein2022, Evered2023, Muniz2024, Finkelstein2024, Tao2025}, utilizing the Rydberg blockade effect for entangling two-, and even multi-body interactions~\cite{Jaksch2000, Isenhower2010, Goerz2014, Levine2019, janduraTimeOptimalTwoThreeQubit2022, Fromonteil2023, Kazemi2025, Ginzel2026}.
Furthermore, the versatility to create arbitrary qubit arrangements~\cite{Barredo2016, Barredo2017, browaeysManybodyPhysicsIndividually2020} and to coherently shuttle atoms during experiments~\cite{Bluvstein2022, Bluvstein2023} allows for an efficient implementation of long-range qubit interactions.
These features have positioned neutral-atom QPUs as a prime platform for quantum error correction (QEC) experiments, with demonstrations of planar and non-planar QEC codes~\cite{Bluvstein2023, Reichardt2024, Bluvstein2025}, transversal logical gates~\cite{Cain2024}, and non-Clifford logical gates through teleportation of magic resource states~\cite{Rodriguez2025, Bluvstein2025}.

In neutral-atom QPUs, single atoms are trapped in a flexible arrangement of optical tweezers or, alternatively, in an optical lattice.
Qubit states are encoded in two atomic low-energy levels, and laser light is used to coherently manipulate and read out the qubits~\cite{Saffman2010, browaeysManybodyPhysicsIndividually2020}.
Entanglement between two or multiple atoms is usually generated through the Rydberg blockade effect, which relies on the strong van der Waals (vdW) interaction between two atoms when they are simultaneously excited to high-energy Rydberg levels~\cite{Saffman2010, browaeysManybodyPhysicsIndividually2020}. 
This interaction mechanism can natively realize CZ gates~\cite{Bluvstein2022, Muniz2024, Finkelstein2024, Tao2025, janduraTimeOptimalTwoThreeQubit2022, Fromonteil2023}, multi control C$^k$Z gates~\cite{Levine2019, Pelegri2022}, and other diagonal multi qubit gates~\cite{Kazemi2025, Locher2025}.

However, the vdW interaction is not the only possible interaction between Rydberg states--nor is it the most fundamental one: It arises as a second-order effect of the underlying dipole-dipole interaction (DDI) between atoms in Rydberg states.
The large orbital size of the electron wave function in Rydberg states leads to substantial atomic dipole moments and, consequently, strong dipole-dipole interactions between atoms in Rydberg states~\cite{Saffman2010}. 
This interaction has been utilized in quantum simulation experiments to explore, for example, coherent spin exchange~\cite{Barredo2015}, continuous symmetry breaking~\cite{Chen2023, Sbierski2023}, and symmetry-protected topological states~\cite{DeLeseleuc2018SSH, yue2025}.
Ultrafast quantum operations~\cite{Chew2022}, CPhase gates~\cite{Goerz2011}, and fast two-qubit CZ gates~\cite{Giudici2025} have also been proposed using the exchange interaction.
Furthermore, entangling gates between individually trapped polar molecules have been realized using such a dipolar exchange interaction~\cite{Bao2023PolarMolecules, Holland2023PolarMolecules, Bergonzoni2025iswapPolarMolecules}.

\begin{figure*}[t]
    \centering
    \includegraphics[width = 1\textwidth]{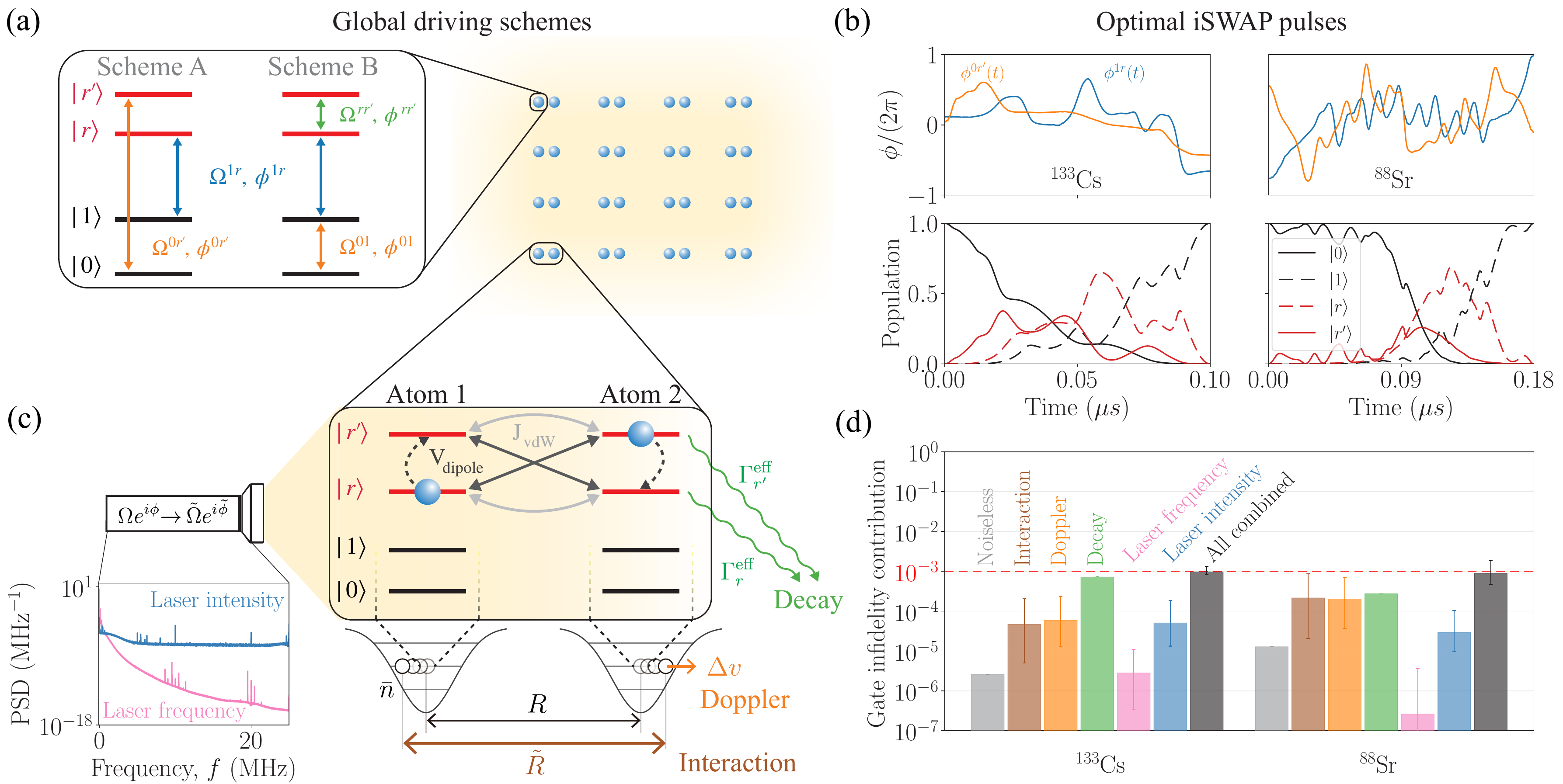}
    \caption{Overview of implementing native iSWAP and exchange gates with Rydberg atoms. 
    (a) We consider pairs of neutral atoms, each with two low-energy (meta)stable qubit states $\ket{0}, \ket{1}$ and two high-energy Rydberg states $\ket{r}, \ket{r'}$, which are dipole coupled and undergo the exchange interaction [Eq.~\eqref{eq:h_exchange}]. The driving is performed with global laser/microwave beams using two different schemes A/B.
    (b) We use optimal control including pulse smoothing techniques to obtain high-fidelity pulse protocols for the iSWAP gate for the two atomic species $^{133}$Cs and $^{88}$Sr. Depicted are resulting pulse profiles for scheme A and phase driving (top panels) and the populations of the energy levels of ``Atom 1'' during the pulse assuming $\ket{0}_1\bigotimes\ket{1}_2$ as the initial pair state (bottom panels).
    (c), (d) Based on an experimentally realistic neutral-atom specific noise model that contains detrimental vdW interactions, atomic motion, Rydberg decay, and laser frequency and intensity fluctuations modeled from PSDs [panel (c)], we evaluate in panel (d) the performance of the optimal pulses shown in panel (b) when subject to each noise source, demonstrating that the pulses achieve overall iSWAP gate fidelities of $99.9\%$ for both $^{133}$Cs and $^{88}$Sr atoms.
    }
    \label{fig:fig1_schemes}
\end{figure*}

In this paper, we utilize the dipole-dipole exchange interaction to realize a native iSWAP gate (and more generally parametrized exchange gates) between two qubits encoded in low-energy states of neutral atoms.
The basic idea is to coherently map populations from the qubit manifold to a pair of dipole-dipole interacting Rydberg states during which the required DDI phase is accumulated, using an optimally controlled pulse protocol.
Similar schemes for iSWAP gates with far separated atoms have also been proposed in Ref.~\cite{Bergonzoni2025}.

Implementing native iSWAP gates (or arbitrary-angle exchange gates) on neutral-atom QPUs promises significant advantages.
In particular, the slower decrease of the relevant exchange interaction with atomic distance, compared to the vdW interaction, offers the potential for faster operation speeds for atoms at larger distances and thereby increased qubit connectivity~\cite{Bergonzoni2025}. 
The combination of nondiagonal iSWAP and exchange gates together with the diagonal CZ gate in a single platform can speed up algorithms and reduce gate count and depth~\cite{Abrams2019, Peterson2020, Krizan2024}.
In quantum optimization, arbitrary-angle exchange gates allow for particle-number-conserving driver terms in variational quantum algorithm circuits~\cite{Wang2020}, and to directly encode optimization sum constraints within the parity architecture for quantum optimization, leading to performance benefits~\cite{Drieb-Schoen2023}.
The recently introduced Parity Twine method~\cite{Klaver2024, Dreier2025}--providing, for example, the currently most efficient implementation of the quantum Fourier transformation~\cite{Dreier2025, Aumann2026}--also benefits from iSWAP gates with further reduced gate count and circuit depths.
Beyond these, iSWAP gates are imperative for implementing promising novel quantum error correction ideas, for example, the recently proposed dynamical surface codes~\cite{Eickbusch2025} and directional quantum low-density parity-check (qLDPC) codes~\cite{Geher2025, Gu2026, Nixon2026}.

In this paper, we demonstrate a native, high-fidelity implementation of iSWAP and exchange gates on neutral-atom QPUs. Figure~\ref{fig:fig1_schemes} provides an overview of our approach and results. 
Our protocol leverages direct DDI between two distinct Rydberg states to facilitate the exchange coupling [Figs.~\ref{fig:fig1_schemes}(a) and  \ref{fig:fig1_schemes}(c)]. 
By combining quantum optimal control with pulse-smoothing techniques to ease experimental constraints, we achieve high-fidelity pulses for both  
$^{133}$Cs and $^{88}$Sr while mitigating detrimental vdW interactions, whose relative contribution is strongly different among the two species [Fig.~\ref{fig:fig1_schemes}(b)]. Finally, we subject our protocol to extensive noise modeling, accounting for both atomic and laser noise sources [Fig.~\ref{fig:fig1_schemes}(c)]. We demonstrate that even in the presence of these fluctuations, gate fidelities of $99.9\%$ are robustly attainable for both atomic species [Fig.~\ref{fig:fig1_schemes}(d)].

The paper is organized as follows: 
Sec.~\ref{sec:setup} details the considered neutral-atom QPU setup and introduces the optimal control framework;
Sec.~\ref{sec:iswap_results} presents our optimal control results for the iSWAP gate for different driving schemes in an idealized, vdW- and noise-free setting for different driving schemes and modulation types;
in Sec.~\ref{sec:vdw}, we analyze the impact of the residual vdW interactions on the optimal control results and perform vdW-inclusive  pulse optimization, specifically for the atomic species $^{133}$Cs and $^{88}$Sr; and Sec.~\ref{sec:noise_budget} presents a detailed analysis of the noise budget for the obtained iSWAP optimal-control protocols under both atomic and laser noise sources described by power spectral densities (PSDs).
Finally, we give our conclusions and outlook in Sec.~\ref{sec:conclusion}. 
In the Appendixes, we provide supplemental information about optimal control results for the parametrized exchange gate, comparison with optimized pulses and the conventional two-pulse-based protocol, reliable hardware settings for our work, the calculation of Rydberg decay rates for $^{88}{\rm Sr}$, and the details about our noise modeling.

\section{Setup and optimal control ansatz}\label{sec:setup}

We consider a neutral-atom quantum computing architecture where individual atoms are trapped in optical tweezers, or in an optical lattice, each encoding a single qubit in two long-lived low-energy states $\ket{0}, \ket{1}$.
Additionally, within each atom we consider two distinct, highly excited, and strongly interacting Rydberg states $\ket{r}, \ket{r'}$ [see Fig.~\ref{fig:fig1_schemes}(a)]; concrete examples for the considered atomic states are given in Appendix~\ref{app:hardware_platforms}.

Using two Rydberg states instead of the usual one gives access to interactions beyond the vdW regime, in particular, allowing the realization of a dipolar exchange interaction~\cite{Saffman2010, Barredo2015, Browaeys2016, browaeysManybodyPhysicsIndividually2020}. 
To illustrate this mechanism, consider, as a toy model, two atoms in Rydberg states $\ket{r_1}$ and $\ket{r_2}$, denoted as ``pair'' state $\ket{r_1 r_2}$, and a second pair state $\ket{r'_1 r'_2}$. If these pair states are dipole-dipole coupled, the interaction takes the form
\begin{equation*}
\hat{H}_{\rm{Ryd}} = \Delta_F \ket{r'_1 r'_2}\bra{r'_1 r'_2} + \frac{C_3}{R^3} \left(\ket{r_1 r_2} \bra{r'_1 r'_2} + \mathrm{H.c.}\right),
\end{equation*}
where $\Delta_F = E_{r'_1} + E_{r'_2} - E_{r_1} - E_{r_2}$ is the F\"orster defect (the difference of atomic energies), $C_3$ is the coupling coefficient, and $R$ is the distance between the atoms.
Diagonalizing $\hat{H}_{\rm Ryd}$ reveals two distinct interaction regimes. 
For $\Delta_F \gg C_3/R^3$, the interaction becomes diagonal, approximated by $\hat{H}_{\rm Ryd} \rightarrow C_6/R^6 \ket{r_1 r_2} \bra{r_1 r_2}$, with the vdW coefficient $C_6 \approx C_3^2 / \Delta_F$.
This is the typical regime in which current QPU setups implement CZ gates, where just a single Rydberg state is used~\cite{Bluvstein2022, Muniz2024, Finkelstein2024, Tao2025, janduraTimeOptimalTwoThreeQubit2022, Fromonteil2023}.
In contrast, for $\Delta_F \ll C_3/R^3$, the interaction becomes off diagonal and exchanges the two pair states coherently, $\hat{H}_{\rm Ryd} \rightarrow C_3/R^3 \left( \ket{r_1 r_2}\bra{r'_1 r'_2} + \mathrm{H.c.} \right)$.
This regime can be reached with very small atomic distances for identical Rydberg states on the two atoms, $\ket{r_1}=\ket{r_2}$, or by tuning the system to a Förster resonance with $\Delta_F \approx 0$, for example, using precise electric or microwave fields~\cite{Walker2005, Ravets2014}. 
Förster resonance tuning has recently been employed for neutral-atom gate engineering, but such approaches are highly susceptible to fluctuations of electric and magnetic fields, and atom position variations~\cite{Ashkarin2025ForsterCCPhase, Kurdak2025ForsterMW}.

In this paper, we use yet another way to enter the dipolar exchange regime, without fine-tuning fields and independent of the lattice distance, by making use of two distinct, directly dipole-coupled, Rydberg states $\ket{r}$ and $\ket{r'}$. 
Then, the most resonant transition channel is $\ket{r r'} \leftrightarrow \ket{r' r}$, with $\Delta_F=0$ by construction, leading (in first order) to the exchange interaction with Hamiltonian~\cite{Barredo2015}
\begin{equation}
    \hat{H}_{\rm Ryd} \rightarrow \hat{H}_{\rm exchange} = \frac{C_3}{R^3} \left( \ket{r r'}\bra{r' r} + \rm{h.c}\right),
\label{eq:h_exchange}
\end{equation}
which coherently exchanges the two Rydberg states $\ket{r}$ and $\ket{r'}$ between two atoms, as illustrated in Fig.~\ref{fig:fig1_schemes}(c). 

The Rydberg interaction coefficient $C_3$ depends on the chosen Rydberg states $\ket{r}$ and $\ket{r'}$ and increases strongly with their principal quantum number as $C_3 \propto n^4$.
Note that the diagonal vdW interactions among the Rydberg states would appear as second-order terms in Eq.~\eqref{eq:h_exchange} (arising from couplings to other, off resonant Rydberg pair states $\ket{p, p'}$). In the first part of this paper, we assume an idealized setting and ignore these interactions. We then analyze their effect in detail in Sec.~\ref{sec:vdw}.

To drive transitions between atomic states, we consider a set of distinct lasers and microwaves that globally illuminate the atoms to reduce experimental requirements.
In particular, we examine the two distinct coupling schemes shown in Fig.~\ref{fig:fig1_schemes}(a) that, by necessity, allow population transfer from the qubit subspace to the two distinct Rydberg states. Driving scheme A directly couples both qubit states to the Rydberg states: $\ket{0}$ with $\ket{r'}$, and $\ket{1}$ with $\ket{r}$ 
\begin{subequations} 
\label{eq:drive_hamiltonian}
    \begin{align}
        H_{\rm drive}^{({\rm A})}(t) / \hbar = \sum_{i} \frac{1}{2} \bigg(  &\Omega^{1r}(t) \E^{\I \phi^{1r}(t)} \ket{r}_i\bra{1} \nonumber\\
         + &\Omega^{0r'}(t) \E^{\I \phi^{0r'}(t)} \ket{r'}_i\bra{0}   + \mathrm{H.c.} \bigg)
         \label{eq:drive_hamiltonianA} .
    \end{align}
Driving scheme B, on the other hand, couples only one qubit state to a Rydberg state, $\ket{1}$ with $\ket{r}$, and additionally the qubit states with each other, and the Rydberg states with each other
    \begin{align}
        H_{\rm drive}^{({\rm B})}(t) / \hbar = \sum_{i} \frac{1}{2} \bigg( &\Omega^{01}(t) \E^{\I \phi^{01}(t)} \ket{1}_i \bra{0}  \nonumber\\
         + &\Omega^{1r}(t) \E^{\I \phi^{1r}(t)} \ket{r}_i\bra{1} \nonumber\\
         + &\Omega^{rr'}(t) \E^{\I \phi^{rr'}(t)} \ket{r'}_i\bra{r}  + \mathrm{H.c.} \bigg)
         \label{eq:drive_hamiltonianB}.
     \end{align}
\end{subequations}
While scheme B is theoretically more complex than scheme A, it is often more easily realizable in existing experiments. This is because scheme A relies on multi step transitions that require explicitly addressing intermediate states in practice. In contrast, scheme B utilizes $\ket{0} \leftrightarrow \ket{1}$ and $\ket{1} \leftrightarrow \ket{r}$ drives that are already standard for single-qubit and CZ gates, meaning the implementation can be completed by adding just a single microwave source for the $\ket{r} \leftrightarrow \ket{r'}$ transition.

The driving Hamiltonians [Eq.~\eqref{eq:drive_hamiltonianA} and \eqref{eq:drive_hamiltonianB}] are written in the corresponding rotating frames.
$\Omega^{ab}(t)$ denotes the time-dependent Rabi frequency of a laser/microwave that couples the atomic states $\ket{a}$ and $\ket{b}$, $\phi^{ab}(t)$ denotes its time-dependent phase, and we only consider resonant driving (i.e., zero detuning).
The idealized system Hamiltonian (excluding detrimental vdW interactions) is then the combination of $H_{\rm drive}$ and the exchange interaction $H_{\rm exchange}$:
\begin{equation}
    H_{\rm ideal}(t) = H_{\rm exchange} + H_{\rm drive}(t) \ ,
    \label{eq:full_hamiltonian}
\end{equation}
where $H_{\rm drive}(t)$ denotes $H_{\rm drive}^{({\rm A})}(t)$ or $H_{\rm drive}^{({\rm B})}(t)$ when driving scheme A or B is chosen, respectively. 
Moreover, we define two types of pulses: Rabi-modulated pulses, where only the Rabi frequencies $\Omega^{ab}(t)$ vary in time, and $\phi^{ab}(t)=0$; and phase-modulated pulses, where only the phases $\phi^{ab}(t)$ vary in time and the Rabi frequencies are set to one constant value $\Omega^{ab}(t) = \Omega_0$.

Our goal is to find optimal pulses that realize an exchange gate of angle $\theta$ between two qubits, given by the unitary (in the qubit subspace)~\cite{Abrams2019}
\begin{equation}
    U_{\rm XY}(\theta) = e^{i \frac{\theta}{2} (X_1 X_2 + Y_1 Y_2)} = 
    \begin{array}{c}
    \begin{matrix}
        \ket{00} & \ket{01} & \ket{10} & \ket{11}
    \end{matrix} \\[2pt]
    \left[\begin{array}{cccc}
        1 & 0 & 0 & 0\\
        0 & \cos\frac{\theta}{2} & i \sin\frac{\theta}{2} & 0\\
        0 & i \sin\frac{\theta}{2} & \cos\frac{\theta}{2} & 0\\
        0 & 0 & 0 & 1
    \end{array}\right]
    \begin{matrix}
        \vphantom{\ket{00}} \\
        \vphantom{\ket{01}} \\
        \vphantom{\ket{10}} \\
        \vphantom{\ket{11}}
    \end{matrix},
    \end{array}
    \label{eq:unitary_gate}
\end{equation}
where $X_i$ and $Y_i$ denote the Pauli-$X$ and Pauli-$Y$ matrices on qubit $i=1,\ 2$.
Importantly, the family of gates $U_{\rm XY}(\theta)$ contains the iSWAP gate for $\theta=\pi$, which is the maximally entangling member of this set. We therefore focus on this case in the remainder of the main text.
Results for other target angles $\theta \neq \pi$ are shown in Appendix~\ref{app:exchange_gate}.

\begin{figure*}[t]
    \centering
    \includegraphics[width = \textwidth]{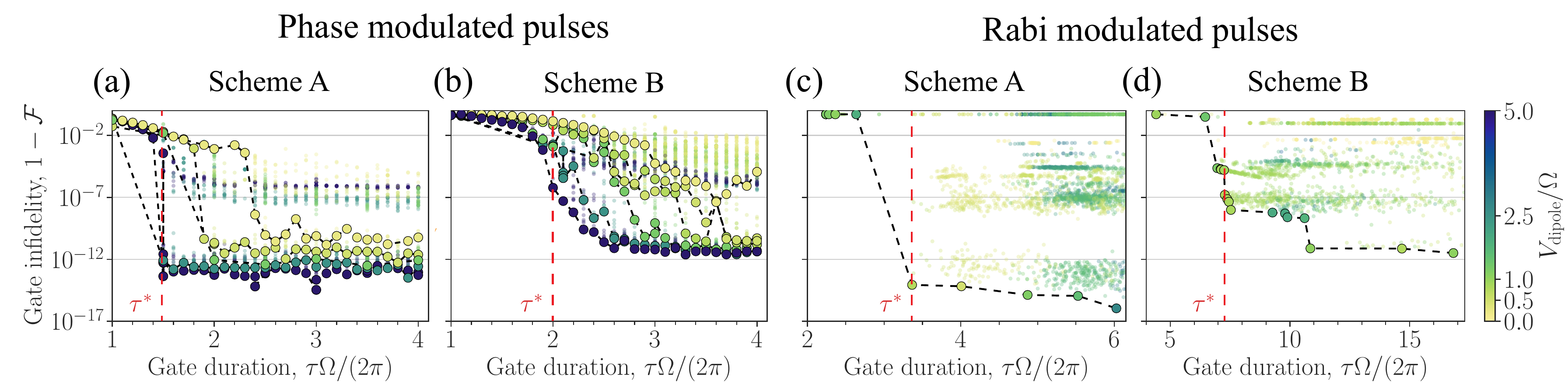}
    \caption{Optimal control results for iSWAP gate. 
    Gate infidelity of different optimal control runs vs unitless gate duration $\tau \Omega$ for phase-modulated pulses for (a) driving scheme A and (b) driving scheme B [cf. Fig.~\ref{fig:fig1_schemes}(a)].
    Same plots for Rabi-modulated pulses for (c) driving scheme A and (d) driving scheme B.
    In all plots, each point corresponds to a single optimal control run with random initialization for a fixed gate duration $\tau$. Highlighted points and black dashed lines are visual guides to the eye tracking the lowest infidelity pulses vs duration.
    Different colors indicate different interaction strengths $V_{\rm dipole}/\Omega$.
    For each scheme and modulation type, we can clearly identify a strong drop of infidelity (where $1-\mathcal{F} = 10^{-6}$ is achieved) at a certain duration $\tau^*$ (red dashed line), which indicates the corresponding iSWAP quantum speed limit for the given setting. 
    For durations $\tau > \tau^*$, we obtain high-fidelity results with infidelities reaching $10^{-10}$ or even below (vdW-free, noiseless).
    }
    \label{fig:pulse_landscapes_all}
\end{figure*}

With the setup considered in this paper, the two atoms naturally undergo the desired exchange interaction within the Rydberg states manifold $\ket{r}, \ket{r'}$, via Eq.~\eqref{eq:h_exchange}. The main remaining challenge in implementing an exchange gate $U_{\rm XY}(\theta)$ is therefore to devise laser and microwave driving schemes that effectively ``transfer'' this interaction into the qubit subspace.
A simple and intuitive approach for realizing $U_{\rm XY}(\theta)$ is a Ramsey-like sequence, or state transfer (ST) protocol, consisting of three steps. First, the atomic populations in the qubit manifold $\{\ket{0}, \ket{1}\}$ are coherently transferred to the Rydberg manifold $\{\ket{r}, \ket{r'}\}$ simultaneously for both atoms. Second, all driving fields are switched off for a designated duration (proportional to $\theta$), allowing the native DDI [Eq.~\eqref{eq:h_exchange}] to generate the target exchange dynamics within the Rydberg manifold. Finally, the first step is inverted to coherently map the populations back from the Rydberg states to the qubit manifold.
However, this simple protocol relies on the assumption that the DDI is negligible during the two state-transfer steps, a condition that is generally difficult to satisfy in practice. 
It would either require fast state transfer (implying $\Omega \gg V_{\rm dipole}$) (see Appendix~\ref{app:ST_protocol} for an analysis on the performance of the ST protocol), or the ability to turn off the DDI during the state transfer steps. 
The latter could be, in principle, achieved using moving atoms~\cite{Bluvstein2022, Shaw2024, Bluvstein2023, Reichardt2024, Bluvstein2025} by performing the state transfer at a large distance between the atoms and bringing them closer during the interaction step.
This is, however, experimentally very demanding, as both qubit and Rydberg states must be simultaneously trapped in movable tweezers, the spatial requirements are substantial, and the required movement speeds would be huge to reach timescales well below the Rydberg lifetimes.

In this paper, we therefore attempt to implement the state transfer and interaction steps concurrently within a time-dependent control pulse and apply \textit{quantum optimal control} methods~\cite{khanejaOptimalControlCoupled2005, janduraTimeOptimalTwoThreeQubit2022, Fromonteil2023} to find high-fidelity protocols for the target gate $U_{\rm XY}(\theta)$, where the time evolution is governed by the Hamiltonian $H(t)$ [Eq.~\eqref{eq:full_hamiltonian}; see Figs.~\ref{fig:fig1_schemes}(a) and \ref{fig:fig1_schemes}(b)].
In particular, we focus on global controls and smooth pulse shapes to relax experimental requirements.
In more detail, for a given duration $\tau$ we try to find time-dependent driving functions $\Omega^{ab}(t)$, $\phi^{ab}(t)$ (for the driving schemes A/B, and either Rabi-modulated or phase-modulated pulse ansätze) such that the Hamiltonian evolution at time $t=\tau$ generates the target unitary $U_{\rm XY}(\theta)$.
Specifically, we request that
\begin{equation}
    U(\tau) = P \tilde{U}(\tau) P = P \, \mathcal{T}\exp\left(-\frac{\I}{\hbar} \int_{0}^{\tau} H(t) \mathrm{d}t\right)\,  P
\end{equation}
is as close as possible to $U_{\rm XY}(\theta)$.
Here, $\tilde{U}(t)$ denotes the Hamiltonian time evolution operator, $P$ the projector into the qubit subspace $\{\ket{0}, \ket{1}\}^{\otimes 2}$, and $\mathcal{T}$ denotes the time-ordering operator.

As a metric for the closeness, we choose the process fidelity/entanglement fidelity
\begin{equation}
    \mathcal{F} = \frac{1}{d^2} \left| \rm{tr}\left( U_{\rm XY}(\theta)^\dagger U(\tau)\right) \right|^2 \, ,
    \label{eq:gate_fidelity}
\end{equation}
between $U(\tau)$ and the target unitary $U_{\rm XY}(\theta)\in \mathbb{C}^{d\times d}$, with $d=2^2$ denoting the dimension of the qubit subspace, and $0 \leq \mathcal{F} \leq 1$. Optimal pulse protocols maximize $\mathcal{F}$, and $\mathcal{F}=1$ corresponds to a perfect implementation of the target unitary.

To find optimal pulses, we consider piecewise constant ansatz functions for the driving terms, similar to the gradient ascent pulse engineering (GRAPE) optimal control method~\cite{khanejaOptimalControlCoupled2005}.
In detail, a function $f(t), 0\leq t\leq \tau$ is divided into $N$ pieces $f_i$, such that $f(t) = f_i$, if $t \in [i\Delta t, (i+1)\Delta t]$ with $\Delta t = \tau/N$.
The values $f_i$ are the optimization parameters, and typically a large number of pieces $N$ is used to receive a good approximation to a smooth function $f(t)$.
We use gradient-based optimization algorithms (BFGS, L-BFGS-B) based on an automatic differentiation (AD) implementation using the software package \textsc{JAX}~\cite{jax2018github} to minimize the gate infidelity $1-\mathcal{F}$.
The advantage of AD is that gradients can be computed automatically even for complicated cost functions and ansätze, enabling the use of highly efficient gradient-based optimization algorithms also for problems where gradients are not known analytically.
We leverage this potential to smooth the pulse wave functions by adding a regularization term to the cost function that penalizes large gradients in the control pulses~\cite{maskaraProgrammableSimulationsMolecules2023}.
In particular, for each control pulse $f(t)$ given as a piecewise function with $N$ pieces $f_i$ we compute its ``smoothness'' cost
\begin{equation}
    C_{\rm smooth}[f] = \sum_{i=0}^{N-1} \left(\frac{f_{i+1} - f_i}{2}\right)^2,
\end{equation}
and the total cost function becomes
\begin{equation}
    C_{\beta} = \left(1 - \mathcal{F}\right) + \beta \sum_{f \in \rm{controls}} C_{\rm smooth}[f] \, .
    \label{eq:cost_function}
\end{equation}
%
Here, $f$ runs over all controls, $\Omega^{ab}(t)$ or $\phi^{ab}(t)$.
Furthermore, the parameter $\beta$ balances smoothness and fidelity to produce smooth optimal pulses ~\cite{maskaraProgrammableSimulationsMolecules2023, Kazemi2025} compatible with limited control bandwidth, a necessary requirement for experimental implementation.
In contrast to feeding already smooth ansatz pulses into the optimizer to achieve smooth high-fidelity gates, the method used here allows for initializing random pulses to avoid getting stuck in low-fidelity local minima. 
Examples of optimal phase-modulated pulse sequences obtained with our optimal control approach are shown in Fig.~\ref{fig:fig1_schemes}(b).

\section{\MakeLowercase{i}SWAP gate results}
\label{sec:iswap_results}

For both driving schemes A and B, and both modulation methods, Rabi and phase modulation, we perform optimal control optimizations for different gate durations $\tau$ and varying DDI strengths $V_{\rm dipole}=C_3/R^3$. For each parameter set, we use multiple (typically between 10 and 20) random initial ansatz pulses, and a smoothing factor $\beta=10^{-3}$.
For phase-modulated pulse protocols, we consider fixed, identical, and constant Rabi frequencies $\Omega^{ab}(t) \equiv \Omega$.
For Rabi modulation, we enforce a lower limit $\Omega^{ab}(t) \geq 0$ in the optimizer but do not constrain the upper limit, and define $\Omega$ as the maximal Rabi frequency obtained among any of the drives throughout the entire pulse; for example, for scheme A, $\Omega = \max\left\{ \max_{t \in [0, \tau]} \Omega^{1r}(t), \max_{t \in [0, \tau]} \Omega^{0r^\prime}(t) \right\}$.

Our results are presented in Fig.~\ref{fig:pulse_landscapes_all}, where each dot represents an optimized pulse.
We find high-fidelity pulses with infidelities $1-\mathcal{F} < 10^{-10}$ for all combinations of driving schemes and modulation methods with the best ones achieving infidelities approaching the limit of double-precision accuracy ($\sim 10^{-16}$).
This demonstrates the potential to realize high-fidelity iSWAP gates in diverse neutral-atom QPU setups with different experimental constraints.
Furthermore, we can clearly identify a strong drop in infidelities after specific pulse durations $\tau^*$, which indicate the quantum speed limit to realize an iSWAP gate for the given setup. In particular, only for durations $\tau \geq \tau^*$ it is possible to realize an iSWAP gate.
The value of $\tau^*$  varies between the driving schemes and modulation types, ranging from $\tau^*\Omega \sim 2 \pi \times 1.5$ for driving scheme A with phase-modulated pulses to $\tau^*\Omega \sim 2 \pi \times 7$ for driving scheme B with Rabi-modulated pulses.
In general, we observe larger speed limits for driving scheme B due to the fact that the state $\ket{r'}$ cannot be populated directly from the qubit subspace, but only via the other Rydberg state $\ket{r}$, unlike in driving scheme A.

The existence of this speed limit arises directly from the basic nature of our protocol, which requires a finite amount of accumulated exchange interaction. Since the interaction strength $V_{\rm dipole}$ is finite, this yields a lower bound on the minimal pulse duration.
Also, finite Rabi frequencies restrict the rate of required transfer of populations from noninteracting qubit states to interacting Rydberg states, and back. However, since our protocols simultaneously perform the required amount of state transfer and interaction accumulation, the observed speed limit $\tau^*$ cannot be simply computed from the interaction strength and used Rabi frequencies. 
The interplay between interaction and Rabi frequency strengths is prominently visible in the results for phase-modulated pulses in Figs.~\ref{fig:pulse_landscapes_all}(a) and \ref{fig:pulse_landscapes_all}(b). One can directly observe that increasing $V_{\rm dipole}/\Omega$ lowers the gate speed limit due to the increased interaction accumulation for larger $V_{\rm dipole}$.  However, there is an upper limit for $V_{\rm dipole}/\Omega$ beyond which no further reduction of the speed limit can be observed anymore. In this regime, the gate speed is limited by the finite Rabi frequencies instead of the interaction strength.

\begin{figure}[t]
    \centering
    \includegraphics[width=1\linewidth]{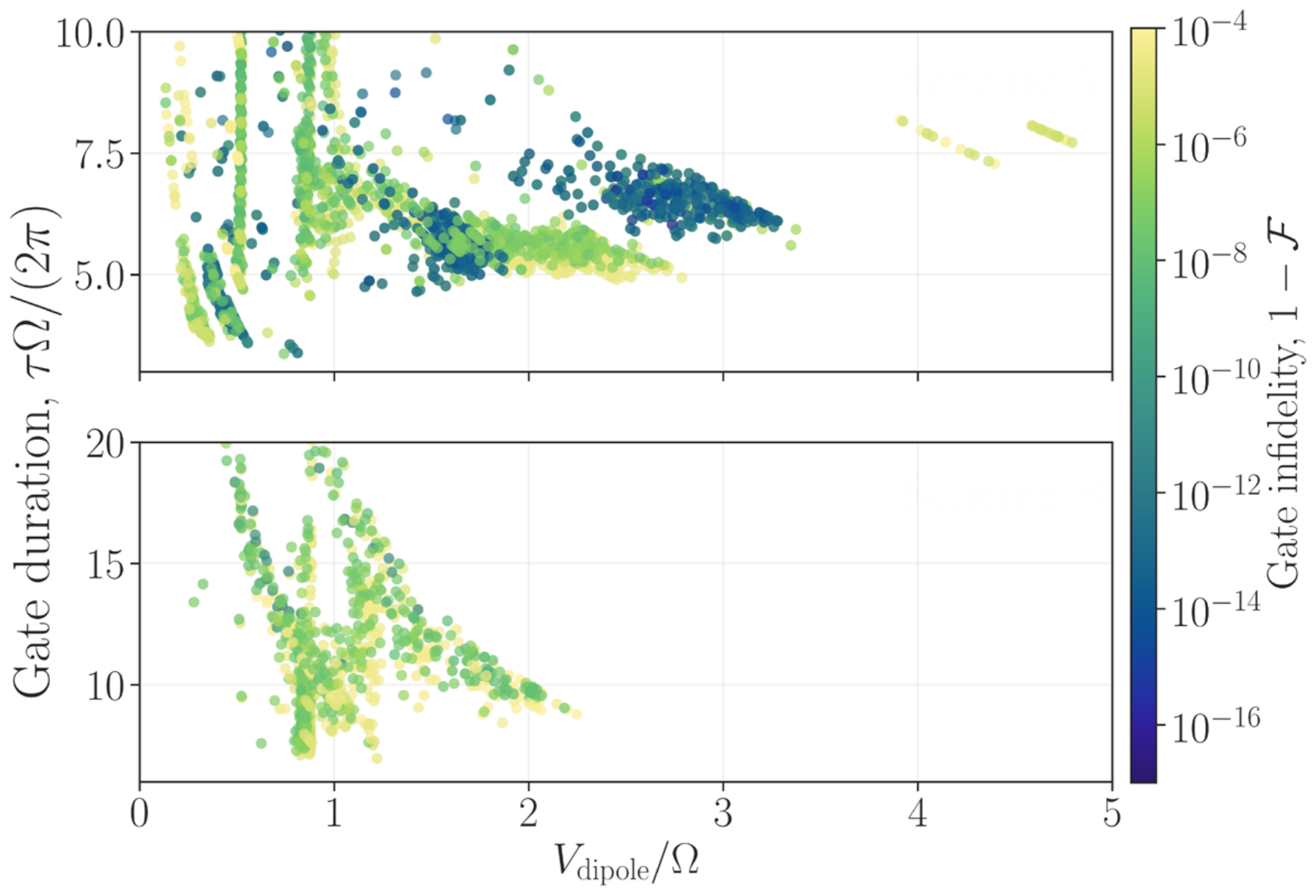}
    \caption{
    Optimal control results for Rabi-modulated drives for (a) scheme A, (b) scheme B with infidelity $1-\mathcal{F} < 10^{-4}$ in the idealized (vdW-free, noiseless) setup.
    We plot the unitless gate duration $\tau\Omega$ as a function of the interaction strength $V_{\rm dipole}/\Omega$. 
    Time-optimal, high-fidelity pulses are obtained for finite interaction strength $V_{\rm dipole} \sim \Omega$. For $V_{\rm dipole} \lesssim \Omega$, the gate duration is limited by finite interaction strength; for $V_{\rm dipole} \gtrsim \Omega$, it is limited by finite Rabi frequency.
    }
    \label{fig:iswap_results_analysis}
\end{figure}

For the Rabi-modulated pulses in Figs.~\ref{fig:pulse_landscapes_all}(c) and \ref{fig:pulse_landscapes_all}(d) this effect is not so clearly visible. The reason is that the maximal Rabi frequency $\Omega$ is automatically optimized for the given interaction strength $V_{\rm dipole}$  during the optimization procedure.
To gain more insight, we plot the optimal control results for the Rabi-modulated pulses in Fig.~\ref{fig:iswap_results_analysis}, where we show the unitless pulse gate durations $\tau \Omega$ as a function of the relative interaction strength $V_{\rm dipole}/\Omega$ for pulses with $1-\mathcal{F} < 10^{-4}$. 
In particular, we find that the optimizer only finds results with $V_{\rm dipole}/\Omega \lesssim 5$ throughout both schemes. Moreover, high-fidelity, short-duration pulses typically obey $V_{\rm dipole} \sim \Omega$.
For $V_{\rm dipole} \lesssim \Omega$, the gate speed is limited by the strength of $V_{\rm dipole}$, which is necessary to generate the required exchange between the Rydberg states. On the other hand, for $V_{\rm dipole} \gtrsim \Omega$ the gate speed is limited by the Rabi frequency $\Omega$, which limits the speed of population transfer between noninteracting qubit states and interacting Rydberg states.

In summary, we find protocols to realize the iSWAP gate with high fidelity in an idealized, noise-free neutral-atom setup for two different driving schemes and both Rabi- and phase-modulated pulses. The speed limit for successful gate implementation is smallest for phase-modulated pulses in driving scheme A reaching $\tau^* \Omega/(2\pi) \approx 1.5$ for optimal values of $2.5 \lesssim V_{\rm dipole}/\Omega \lesssim 5$. Notably, this number is just about $25\%$ larger than the time-optimal CZ gate demonstrated in Ref.~\cite{janduraTimeOptimalTwoThreeQubit2022}.

\section{Residual van der Waals interactions in realistic hardware platforms}
\label{sec:vdw}

The results presented in Sec.~\ref{sec:iswap_results} assumed an idealized model for neutral atoms, where residual vdW interactions between Rydberg states were completely neglected.
However, in real hardware implementations, vdW interactions will not vanish completely and influence the gate performance.
In this section, we analyze their impact on the gate fidelity and discuss strategies to mitigate them.

Because the strength of vdW interactions depends heavily on the atomic species and the specific Rydberg states, we focus on two representative species from the alkali and alkaline-earth families: $^{133}$Cs and $^{88}$Sr atoms.
$^{133}$Cs is a well-established alkali atom in neutral-atom platforms~\cite{Manetsch2024, anand2024dual, graham2022multi, graham2023midcircuit}, where the qubit is usually encoded in the ground state's hyperfine qubit (HFQ) manifold. 
$^{88}$Sr, as one of the alkaline-earth atoms, offers a particularly rich atomic level structure for quantum computing~\cite{Scholl2023, Endres2025Benchmark, Tao2025}, with two distinct, commonly used qubit encodings, the clock qubit (CQ)~\cite{PhysRevLett.91.173005, ludlow2008sr, 5422497, PhysRevLett.103.063001, PhysRevLett.106.210801, PhysRevA.84.052716, PhysRevLett.124.203201} and the fine-structure qubit (FSQ)~\cite{meinert2021quantum, PhysRevLett.132.150605, PhysRevLett.132.150606}.
In Appendix \ref{app:hardware_platforms}, we propose specific implementations of the driving schemes A and B [cf. Fig.~\ref{fig:fig1_schemes}(a)] for both of these atomic species and their common qubit encodings.

\begin{figure}[t]
    \centering
    \includegraphics[width=\columnwidth]{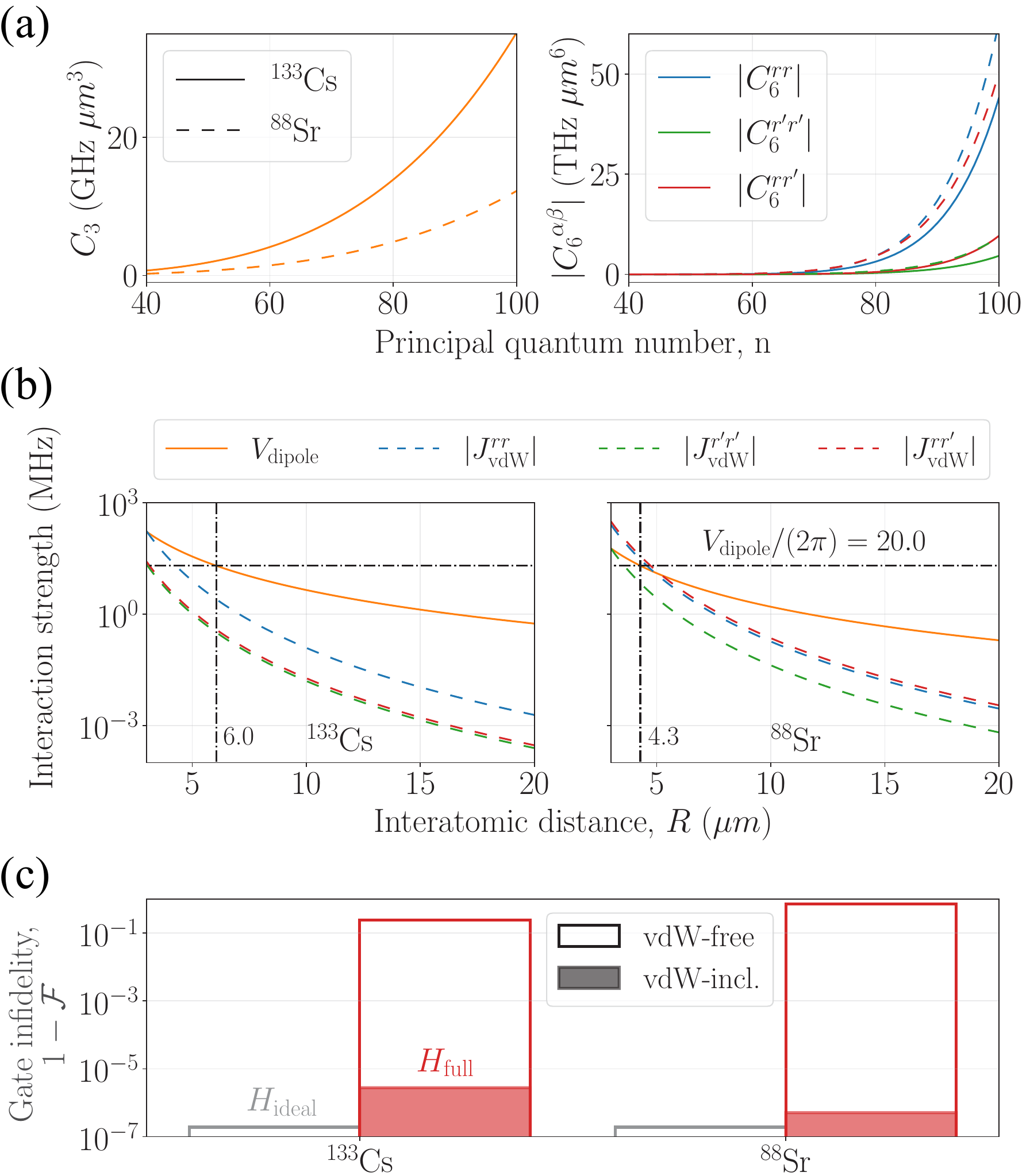}
    \caption
    {van der Waals interactions and effect on gate fidelity. 
    (a) Comparison of $C_3$ (left) and $C_6$ (right) interaction coefficients between $^{133}$Cs and $^{88}$Sr for different principal quantum numbers $n$. 
    $^{133}$Cs clearly shows an advantage over $^{88}$Sr with a larger $C_3$ coefficient and overall smaller $C_6$ 
    coefficients, suggesting that $^{88}$Sr-based gates are more susceptible to vdW interactions.
    (b) Exchange interaction strength $V_{\rm dipole}$ and the vdW interaction strengths $J_{\rm vdW}^{\alpha\beta}$ for each species as a function of the interatomic distance $R$ for $n=61$. The vertical dash-dotted lines represent the interatomic distances for which $V_{\rm dipole} = 2\pi \times 20$~MHz.
    (c) Gate infidelity for pulses optimized with and without the inclusion of vdW interactions. Gray bars denote infidelity evaluated against the idealized Hamiltonian ($H_{\rm{ideal}}$), while red bars show infidelity under the full system Hamiltonian ($H_{\rm{full}}$). Empty bars correspond to vdW-free optimized pulses; filled bars correspond to vdW-inclusive optimized pulses. For both $^{133}\text{Cs}$ and $^{88}\text{Sr}$, pulses optimized without vdW considerations are heavily corrupted when evaluated under $H_{\rm{full}}$, driving infidelities above $10^{-1}$ (empty red bars). However, vdW-inclusive pulse shaping successfully suppresses these errors, recovering high-fidelity gates with $1-\mathcal{F} \lesssim 10^{-5}$ (filled red bars)---even for $^{88}\text{Sr}$, where vdW interactions are comparable in strength to dipole interactions. We assume $\Omega = 2\pi \times 10\text{ MHz}$, $n=61$, and an interatomic distance $R$ defined by the vertical dashed lines in panel (b) corresponding to time-optimal pulses.
    }
    \label{fig:non_vdw_vs_vdw_optimised_pulse_C3_C6_coeffs}
\end{figure}

The inclusion of the vdW interactions adds an additional, time-independent term
\begin{align}
    H_{\rm vdW} = &J_{\rm{vdW}}^{rr} \ket{r r}\bra{r r} + J_{\rm{vdW}}^{r'r'} \ket{r' r'}\bra{r' r'} \nonumber \\
    + &J_{\rm{vdW}}^{rr'} \left( \ket{r r'}\bra{r r'} + \ket{r' r}\bra{r' r}\right) \, ,
     \label{eq:vdw_hamiltonian_noise}
\end{align}
to the system Hamiltonian, where $J_{\rm vdW}^{\alpha\beta} = C_6^{\alpha \beta}/R^6$ and $C_6^{\alpha \beta}$ denote the species-dependent vdW interaction coefficients for the Rydberg pair interaction $\ket{\alpha \beta} \bra{\alpha \beta}$. The full Hamiltonian is then given 
\begin{equation}
    H_{\rm full}(t) = H_{\rm exchange} + H_{\rm vdW} + H_{\rm drive}(t) \, .
    \label{eq:vdW_full_hamiltonian}
\end{equation}
The additional, diagonal terms in the Hamiltonian lead to additional detrimental phases acquired throughout the optimal control protocol while atoms are excited to the Rydberg states and, thus, diminish the gate quality. 
The strength of this effect depends on two aspects: (1) the strength of the vdW interaction coefficients $C_6^{\alpha \beta}$ vs the dipole interaction coefficient $C_3$; and (2) the interatomic distance $R$ because of the different scalings of the exchange interaction ($\propto\!\!R^{-3}$) and vdW interactions ($\propto\!\!R^{-6}$). 
In Fig.~\ref{fig:non_vdw_vs_vdw_optimised_pulse_C3_C6_coeffs}(a), we analyze aspect (1) for $^{133}$Cs and $^{88}$Sr as a function of different principal quantum numbers $n$, with data obtained from the software \textit{pairinteraction}~\cite{mögerle2026accuratemodelingrydbergatoms}\footnote{We compute the $C_3$ and $C_6$ coefficients independently for each $n$ via perturbation theory of the effective Hamiltonian in the subspace of $\ket{r}$ and $\ket{r^{\prime}}$ with approximately $100,000$ basis pair states.}.
Note that $C_6^{r'r'}$ for $^{133}$Cs, and both $C_6^{r'r'}$ and $C_6^{rr'}$ for $^{88}$Sr, carry a negative sign; for ease of comparison between species, we plot their absolute values. 
We clearly observe that the $C_3$ coefficients are much larger for $^{133}$Cs than for $^{88}$Sr, while the detrimental vdW coefficients are smaller in $^{133}$Cs than in $^{88}$Sr, especially for $C_6^{rr'}$.
The latter is fundamentally important because the pair states $\ket{r r'}$ and $\ket{r' r}$ necessarily need to be excited for some duration to successfully realize the desired gate. 
Therefore, we expect that the vdW interactions are less problematic for $^{133}$Cs than for $^{88}$Sr and that the range of interatomic distances, where vdW interactions are perturbative, is larger.
Another important aspect to consider is the different scaling of the coefficients with the principal quantum number $n$. While $C_3 \propto n^4$, the vdW coefficients scale much faster, $C_6^{\alpha \beta} \propto n^{11}$. 
Thus, for a fixed interatomic distance, the relative effect of vdW interactions quickly increases with $n$.

In Fig.~\ref{fig:non_vdw_vs_vdw_optimised_pulse_C3_C6_coeffs}(b), we analyze aspect (2) by plotting the exchange interaction strength $V_{\rm dipole}$ and the vdW interaction strengths $J_{\rm vdW}^{\alpha\beta}$ as a function of interatomic distance $R$ for a fixed principal quantum number $n=61$. 
We clearly observe that, due to their faster $R^{-6}$ scaling, vdW interactions become quickly important at short distances, with the largest ones reaching the exchange interaction strength at $R\approx 3 ~\si{\micro\meter}$ ($R\approx 5~ \si{\micro\meter})$ for $^{133}$Cs ($^{88}$Sr) for the chosen $n$.
At large distances, however, the slow decay of the exchange interactions compared to the vdW interactions makes the former still retain reasonably large values, allowing for long-range gates~\cite{Bergonzoni2025}, and the vdW interactions become more and more perturbative.
Interestingly, the described features are qualitatively shared by $^{87}$Rb and $^{174}$Yb, which behave similarly to their alkali and alkaline-earth relatives $^{133}$Cs and $^{88}$Sr, respectively (not shown).

We continue by computing the impact of the vdW interactions on the previously obtained optimal iSWAP pulse sequences for the two species.
For that, we choose a pulse close to the time-optimal limit with $\tau\Omega = 2\pi \times 0.25$, we fix the principal quantum number $n=61$, the Rabi frequency $\Omega/(2\pi) = 10$~MHz, and the dipole interaction strength $V_{\rm dipole}/(2\pi) = 20$~MHz. Because of the different $C_3$ coefficients, this leads to different interatomic distances and different relative impact of the vdW interactions for the two considered atomic species [see dashed lines in Fig.~\ref{fig:non_vdw_vs_vdw_optimised_pulse_C3_C6_coeffs}(b)].
We then evaluate the gate infidelity of this pulse under the time evolution with the full Hamiltonian $H_{\rm full}(t)$ including the vdW interactions for the two different species. 
The result is shown as empty bars in Fig.~\ref{fig:non_vdw_vs_vdw_optimised_pulse_C3_C6_coeffs}(c), and we observe that the vdW interactions strongly increase the gate infidelity beyond experimentally interesting levels.
This happens even for $^{133}$Cs, where the vdW interaction strengths are more than an order of magnitude smaller than $V_{\rm dipole}$, and even more so for $^{88}$Sr where the vdW interactions are similar to $V_{\rm dipole}$, for the given parameters [cf. Fig.~\ref{fig:non_vdw_vs_vdw_optimised_pulse_C3_C6_coeffs}(b)].

Crucially, however, the coherent nature of the detrimental vdW interactions makes it amenable to compensation through pulse shaping.
By explicitly considering $H_{\rm full}(t)$ during the optimization, the optimizer can adapt the pulse to counteract or utilize the additional phases accumulated from vdW interactions during the gate. 
As shown by the filled bars in Fig.~\ref{fig:non_vdw_vs_vdw_optimised_pulse_C3_C6_coeffs}(c), this vdW-inclusive optimization recovers infidelities below $10^{-5}$, demonstrating that vdW interactions, despite their severity, can be effectively absorbed into the pulse design.

\subsection{vdW-inclusive pulse optimization}

\begin{figure*}[t]
    \centering
    \includegraphics[width=\textwidth]{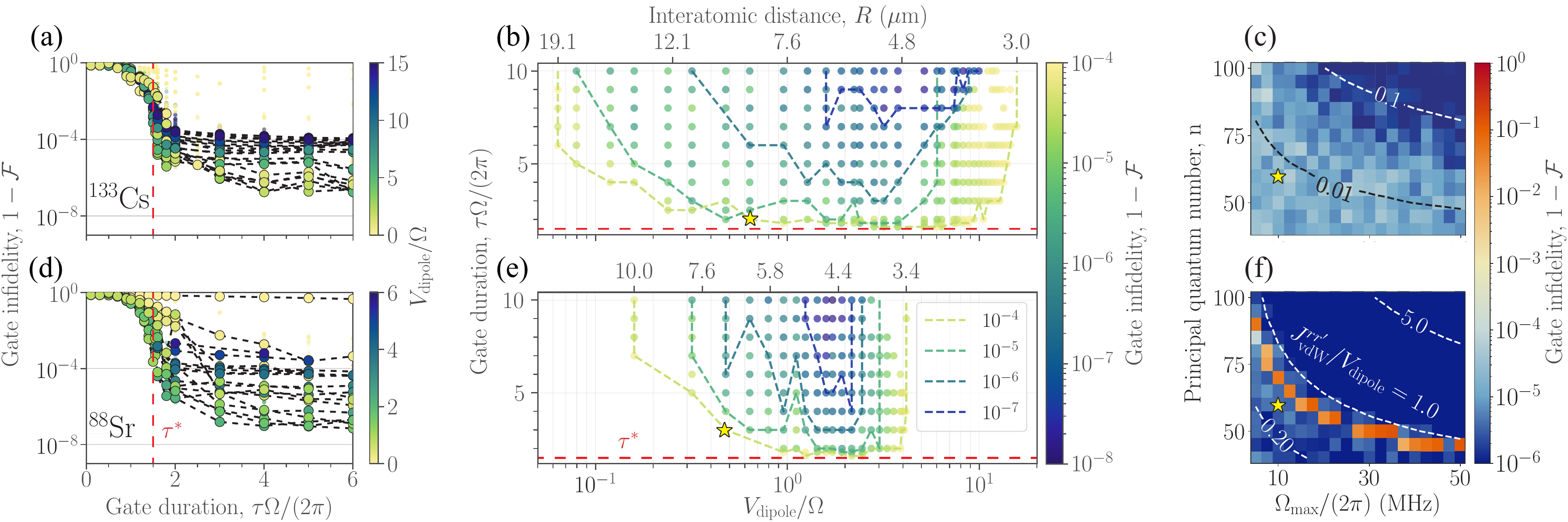}
    \caption{vdW-inclusive optimal control results for the iSWAP gate (scheme A, phase-modulated pulses) for $^{133}$Cs [panels (a)-(c)] and $^{88}$Sr [panels (d)-(f)] based QPUs, respectively. 
    (a), (d) Gate infidelity vs unitless gate duration $\tau \Omega$ for fixed $\Omega=2\pi\times10$~MHz and $n=61$. We indicate the $H_{\rm ideal}$ speed limit $\tau^*$ and find that the $H_{\rm full}$ speed limit remains mostly unchanged relative to $\tau^*$ [cf. Fig.~\ref{fig:pulse_landscapes_all}(a)]. 
    (b), (e) Subset of high-fidelity pulses ($1-\mathcal{F} < 10^{-4}$) with unitless duration $\tau \Omega$ plotted against the ratio between the DDI and the Rabi frequency $V_{\rm dipole}/\Omega$. 
    High-fidelity speed-limit pulses can be found within an extended range of $V_{\rm dipole}/\Omega$ as seen by the flat, bottom boundaries. For $^{133}$Cs, (b) we find a considerably wider range than for $^{88}$Sr, (e) which results in an increased range of interatomic distances for which fast, high-fidelity gates can be performed [see top axis in panels (b) and (e)]. 
    (c), (f) Gate infidelity of optimal pulses as a function of Rabi frequency $\Omega$ and principal quantum number $n$ for fixed $\tau\Omega$ and $V_{\rm dipole}/\Omega$ chosen from panels (b) and (e) such that the time-optimal condition is maintained [cf. yellow stars in panels (b) and (e)]. Dashed white and black curves indicate constants of the interaction ratio $J_{\rm vdW}^{rr'}/V_{\rm dipole}$. See text for details.
    }
    \label{fig:iswap_Sr_Cs_results_analysis}
\end{figure*}

We perform vdW-inclusive optimizations for $\Omega = 2\pi \times 10$~MHz and $n=61$ over a large space of gate durations $\tau \in \left[0.01,\ 1.0 \right]\ \si{\micro\second}$, and DDI strengths $V_{\rm dipole}/(2\pi) \in \left[0.6,\ 160 \right]$ MHz for both $^{133}$Cs and $^{88}$Sr.
Figures~\ref{fig:iswap_Sr_Cs_results_analysis}(a) and \ref{fig:iswap_Sr_Cs_results_analysis}(d) show the fidelity of the optimal control results vs gate duration for $^{133}$Cs and $^{88}$Sr, respectively.
Notably, above some duration $\tau>\tau^*_{\rm vdW}$ we observe a drop of the iSWAP gate infidelity to values around $10^{-6}$ or even below for a large range of interaction strengths $V_{\rm dipole}$. 
Importantly, this drop happens at gate durations similar or only slightly above the fundamental speed limit of the vdW-free optimal pulses of $\tau^*\Omega \sim 2\pi \times 1.5$ [see red dashed lines in Fig.~\ref{fig:iswap_Sr_Cs_results_analysis}(a) and \ref{fig:iswap_Sr_Cs_results_analysis}(d) and cf. Fig.~\ref{fig:pulse_landscapes_all}(a)], showing that the inclusion of vdW interactions can mostly preserve the speed limit for both considered atomic species for the given parameters.

In Figs.~\ref{fig:iswap_Sr_Cs_results_analysis}(b) and \ref{fig:iswap_Sr_Cs_results_analysis}(e) we plot the unitless gate duration $\tau\Omega$ of all obtained vdW-inclusive optimal control results with infidelity $1-\mathcal{F} < 10^{-4}$ against the ratio $V_{\rm dipole}/\Omega$. For both considered atomic species, we find a rather extended flat bottom region, showing that pulses close to the speed limit $\tau^*$ can be reached for interaction strengths $V_{\rm dipole}/\Omega \in [0.5, 10] \, ([1, 3])$ for $^{133}$Cs ($^{88}$Sr) at $\Omega = 2\pi \times 10$~MHz.
This extended range is a feature of phase-modulated pulses and is in stark contrast to Rabi-modulated pulses, where we observed rather narrow optimal regimes for $V_{\rm dipole}/\Omega$ for the vdW-free case (cf. Fig.~\ref{fig:iswap_results_analysis}]).
Importantly, for fixed principal quantum number, here $n=61$, this also defines a range of interatomic distances $R$ for which the iSWAP gate can be implemented with high fidelity and close to the speed limit under the presence of vdW interactions, as shown by the top axes labels in Figs.~\ref{fig:iswap_Sr_Cs_results_analysis}(b) and \ref{fig:iswap_Sr_Cs_results_analysis}(e).
For the given parameters, a gate close to the speed limit with high fidelity can be obtained for $3.5 ~\si{\micro\meter} \lesssim R \lesssim 12 ~\si{\micro\meter}$ ($3.5 ~\si{\micro\meter} \lesssim R \lesssim 7.5 ~\si{\micro\meter}$) for $^{133}$Cs ($^{88}$Sr).
We attribute the difference to the different strengths of the $C_3$ and the $C_6^{\alpha \beta}$ coefficients among the species, where the reduced ratios of $C_3/|C_6^{\alpha\beta}|$ for $^{88}$Sr restrict high-fidelity gates to a narrower regime of interaction strengths and, therefore, interatomic distances.

Since the vdW-inclusive optimization results depend on specific experimental parameters--the atomic species, the principal quantum number $n$, the interaction strength $V_{\rm dipole}$ (which together with the two previous parameters fixes the interatomic distance $R$), the Rabi frequency $\Omega$, and the pulse duration $\tau$--it is essential to evaluate the optimizer's performance across these varying regimes. 
To simplify the large parameter space $\left\{{\rm species},\ n,\ V_{\rm dipole},\ \Omega,\ \tau \right\}$ we fix the unitless quantities $\tau \Omega$ and $V_{\rm dipole}/ \Omega$ individually for both species based on the results in Figs.~\ref{fig:iswap_Sr_Cs_results_analysis}(b) and \ref{fig:iswap_Sr_Cs_results_analysis}(e).
More specifically, from the given results, we jointly minimize both unitless quantities while maintaining small infidelities $1-\mathcal{F} < 10^{-4}$; minimizing $\tau \Omega$ is crucial to keep gate operations as fast as possible, thereby reducing fundamental decoherence from Rydberg state decay; minimizing $V_{\rm dipole}/ \Omega$ reduces decoherence from interaction fluctuations due to atom position noise, another major noise source (see Sec.~\ref{sec:noise_budget} for a detailed discussion of noise effects).
In particular, we choose the parameters $\tau \Omega = 2.0 \, (3.0)$ and $V_{\rm dipole}/ \Omega = 0.64 \, (0.47)$ for $^{133}$Cs ($^{88}$Sr) [indicated in Fig.~\ref{fig:iswap_Sr_Cs_results_analysis}(b) and \ref{fig:iswap_Sr_Cs_results_analysis}(e) with a yellow star].
To verify this choice, we also analyzed alternative sets of parameters along the 
speed-limit boundary by shifting the trade-off between $\tau\Omega$ and 
$V_{\rm dipole}/\Omega$. Their sensitivity to Rydberg decay and interaction 
fluctuations was benchmarked using the same framework presented in 
Sec.~\ref{sec:noise_budget}, confirming that the selected values 
represent the optimal noise-resilient configuration.

Using these parameters, we find optimal pulses for the iSWAP gate across a range of experimentally possible principal quantum numbers $n \in \left[40,\ 100 \right]$ and Rabi frequencies $\Omega/(2\pi) \in \left[ 0.1,\ 50\right]$.
The resulting infidelities are plotted in Figs.~\ref{fig:iswap_Sr_Cs_results_analysis}(c) and \ref{fig:iswap_Sr_Cs_results_analysis}(f) for $^{133}$Cs and $^{88}$Sr, respectively.
For $^{133}$Cs, we obtain high-fidelity pulses, $1-\mathcal{F} \lesssim 10^{-4}$, across the whole parameter space, and even reach infidelities of $1-\mathcal{F} \lesssim 10^{-6}$ for large $n$ and $\Omega$. 
Since $V_{\rm dipole}/\Omega$ is fixed throughout this plot, and due to the scalings of the dipole and vdW interaction strengths with $n$, this is surprising, because in this regime the relative vdW interaction contributions $J_{\rm vdW}^{\alpha \beta}/V_{\rm dipole}$ are larger than for small $n$ and $\Omega$, as indicated by the contour lines of constant $J_{\rm vdW}^{\alpha \beta}/V_{\rm dipole}$ in Fig.~\ref{fig:iswap_Sr_Cs_results_analysis}(c).
On the one hand, this demonstrates once more the effectiveness of the vdW-inclusive optimal control approach. On the other hand, it suggests that the presence of vdW interactions does not necessarily degrade gate performance and may, in some regions of parameter space, be exploited by the optimization to assist the implementation of the exchange gate.

For $^{88}$Sr, where vdW interactions are generally large compared to $^{133}$Cs [see contour lines of $J_{\rm vdW}^{\alpha\beta}/V_{\rm dipole}$ in Figs.~\ref{fig:non_vdw_vs_vdw_optimised_pulse_C3_C6_coeffs}(a), \ref{fig:non_vdw_vs_vdw_optimised_pulse_C3_C6_coeffs}(b), Figs.~\ref{fig:iswap_Sr_Cs_results_analysis}(c) and \ref{fig:iswap_Sr_Cs_results_analysis}(f)] we find pulses with very low infidelities $1-\mathcal{F} \lesssim 10^{-5}$ almost throughout the entire considered parameter space of $n$ and $\Omega$ [see Fig.~\ref{fig:iswap_results_analysis}(f)].
However, we also observe a curve of parameters for which the optimizer struggles to find any high-fidelity pulses. Interestingly, this curve follows closely the shape of parameters for which $|J^{rr'}|/V_{\rm dipole} \sim 1$.
We leave a more detailed analysis of this interesting feature to future work.
These results clearly demonstrate that high-fidelity iSWAP and exchange gates can be implemented even when vdW interactions are (in absolute values) larger than the exchange interactions in the system, and do not necessarily require the vdW interactions to be perturbatively small.

\section{\MakeLowercase{i}SWAP noise budget}
\label{sec:noise_budget}

\begin{figure*}[t]
    \centering
    \includegraphics[width=\textwidth]{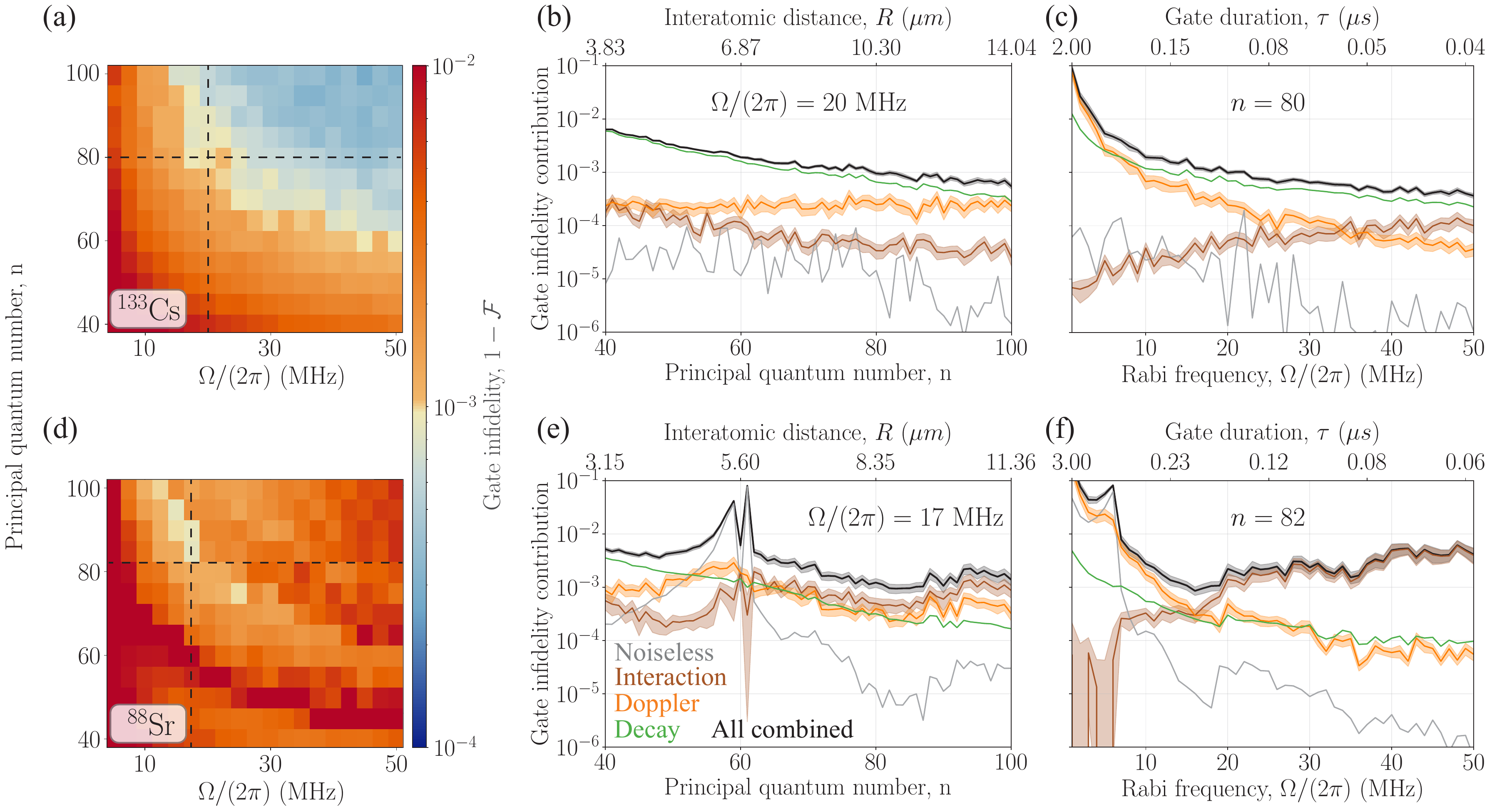}
    \caption{
    Noise sensitivity of vdW-inclusive optimized pulses for $^{133}$Cs [panels (a)-(c)] and $^{88}$Sr [panels (d)-(f)] atoms. 
    (a), (d) Combined gate infidelity from all atomic noise sources (interaction, Doppler, decay noise) vs Rabi frequency $\Omega$ and principal quantum number $n$.
    $^{133}$Cs shows an extended region with $1-\mathcal{F} \leq 10^{-3}$ when $\Omega$ and $n$ are increased [panel (a)], while this region is very limited in parameter space for $^{88}$Sr [panel (d)].
    For $^{133}$Cs, the noise analysis is based on the optimal control results presented in Fig.~\ref{fig:iswap_Sr_Cs_results_analysis}(c), while for $^{88}$Sr we base the analysis on noise-aware pulse optimization results [assuming identical parameters as in Fig.~\ref{fig:iswap_results_analysis}(f)] to reduce the contribution of interaction noise (see main text for details).
    (b), (e) Contributions of the individual noise sources to the gate infidelity vs principal quantum number $n$ for fixed $\Omega$ [cf. vertical dashed lines in panels (a) and (d)]. 
    (c), (f) Contributions of the individual noise sources to the gate infidelity vs Rabi frequency $\Omega$ for fixed $n$ [cf. horizontal dashed lines in panels (a) and (d)]. 
    See main text for details.
    }
    \label{fig:iswap_Sr_Cs_noise_analysis}
\end{figure*}

In the previous section, we demonstrated high-fidelity iSWAP pulses using vdW-inclusive optimization over a large parameter space assuming a noiseless system. 
In actual hardware, however, environmental noise processes~\cite{deLeseleuc2018} degrade the system dynamics and reduce gate fidelities.
In this section, we consider three types of noise: motion of the atoms, atomic decay, and intensity/frequency fluctuations from the driving fields [see Fig.~\ref{fig:fig1_schemes}(c)].

We model the noise induced by atomic motion with the semiclassical ``frozen gas'' approximation, by assuming that the atoms are frozen during the gate operation, and its effect can be sampled on a shot-to-shot basis. 
Within this approximation, atomic motion introduces two distinct noise mechanisms.
The first, denoted as ``interaction noise'', randomly samples the position of both atoms within their traps and, therefore, alters the atomic distance $R$ by a small amount $\Delta R \ll R$, which directly affects both the DDI and vdW interaction strengths $V_{\rm dipole} \rightarrow \tilde{V}_{\rm dipole} = C_{3} / (R + \Delta R)^{3}$, $J_{\rm vdW}^{\alpha\beta} \rightarrow \tilde{J}_{\rm vdW}^{\alpha\beta} = C_{6}^{\alpha\beta} / (R + \Delta R)^{6}$.
Second, detrimental Doppler shifts of the lasers are induced from finite atom velocities within their traps.
We denote this effect as ``Doppler noise'' and model it with additional detuning terms
\begin{align}
    \frac{\tilde{H}_{\rm Doppler}}{\hbar} = - \sum_{i} \bigg(\Delta^{1} \ket{1}_i\bra{1}_i &+ \Delta^{r} \ket{r}_i\bra{r}_i \nonumber \\
    &+ \Delta^{r'} \ket{r'}_i\bra{r'}_i \bigg) \ , 
     \label{eq:det_hamiltonian_noise}
\end{align}
added to the Hamiltonian $H_{\rm full}(t)$, where $\Delta^a$ denotes the Doppler-induced detuning for atomic state $\ket{a}$.

Atomic decay from the Rydberg states, denoted as ``decay noise'' in the following, is implemented through a non-Hermitian Hamiltonian~\cite{Pagano2022}, 
\begin{align} \label{Decay_ham}
    \frac{\tilde{H}_{\rm decay}}{\hbar} = - \frac{i}{2} \sum_{i} \bigg(\Gamma^{\text{eff}}_{r} \ket{r}_i\bra{r}_i +  \Gamma^{\text{eff}}_{r^\prime} \ket{r'}_i\bra{r'}_i \bigg) \ ,
\end{align}
where $\Gamma_{r}^{\text{eff}}$ and $\Gamma_{r'}^{\text{eff}}$ represent the effective decay rates of the Rydberg states $\ket{r}$ and $\ket{r'}$. In this noise analysis, we neglect blackbody radiation (BBR) when computing $\Gamma^{\text{eff}}_r$ and $\Gamma^{\text{eff}}_{r'}$, an approximation that holds well for modern cryogenic setups~\cite{Schymik2021, Pichard2024}. For $^{88}$Sr, the relevant values of radiative decay rates have been evaluated using methods based on the quantum defect theory approach reported in Ref.~\cite{kostelecky1985analytical} (see Appendix~\ref{sec:App/FSQubitDecay} for details). For $^{133}$Cs, we used the software pairinteraction~\cite{mögerle2026accuratemodelingrydbergatoms}.

We describe laser intensity and frequency fluctuations by PSDs [see Fig.~\ref{fig:fig1_schemes}(c)] and, for simplicity, we assume identical PSDs for all driving fields in our system. 
Laser intensity noise alters the Rabi frequencies $\Omega^{ab}(t) \rightarrow \tilde\Omega^{ab}(t)$, while laser frequency noise changes $\phi^{ab}(t) \rightarrow \tilde\phi^{ab}(t)$. 
The modified $\tilde\Omega^{ab}(t)$ and $\tilde\phi^{ab}(t)$ can be obtained from the corresponding PSD using a Monte Carlo sampling approach~\cite{deLeseleuc2018, Jiang2022} (details are provided in Appendix~\ref{App:phase}).

Combining all these terms, the full noisy Hamiltonian becomes
\begin{equation}
    H_{\rm noisy}(t) = \tilde{H}_{\rm full}(t) + \tilde{H}_{\rm Doppler} + \tilde{H}_{\rm decay} \ ,
\end{equation}
where $\tilde{H}_{\rm full}(t)$ corresponds to the noise-free Hamiltonian  [Eq.~\eqref{eq:vdW_full_hamiltonian}], with the replacements $V_{\rm dipole} \rightarrow \tilde{V}_{\rm dipole}$, $J_{\rm vdW}^{\alpha\beta} \rightarrow \tilde{J}_{\rm vdW}^{\alpha\beta}$ and $\Omega^{ab}(t) \rightarrow \tilde\Omega^{ab}(t), \phi^{ab}(t) \rightarrow \tilde\phi^{ab}(t)$.
Further details about the noise modeling are provided in Appendix~\ref{sec:A/noise_model}.

In Sec.~\ref{sec:atomic_noise}, we begin with a detailed analysis of the impact of \textit{atomic noise}--i.e. interaction and Doppler noise from atomic motion and decay noise--on the iSWAP fidelities reported in Sec.~\ref{sec:vdw}. 
We then consider laser-induced noise in Sec.~\ref{sec:laser_noise}.

\subsection{Atomic noise}
\label{sec:atomic_noise}

Figure~\ref{fig:iswap_Sr_Cs_noise_analysis} analyses the resulting gate infidelities of the vdW-inclusive optimal pulses shown before in Figs.~\ref{fig:iswap_Sr_Cs_results_analysis}(c) and \ref{fig:iswap_Sr_Cs_results_analysis}(f) when atomic noise (interaction, Doppler, and decay noise) is applied on those. 
The overall infidelity as a function of the parameters $\Omega$ and $n$ is shown in Figs.~\ref{fig:iswap_Sr_Cs_noise_analysis}(a) and \ref{fig:iswap_Sr_Cs_noise_analysis}(d) for $^{133}$Cs and $^{88}$Sr, respectively.
Note that, as briefly mentioned above, the fixed dimensionless parameters $\tau\Omega$ and $V_{\rm{dipole}}/\Omega$ are correlated with atomic decay and interaction noise, and were chosen such that their effects are minimized. In short, a longer gate duration increases the sensitivity to Rydberg state decay, while a stronger interaction strength for fixed $n$--shorter interatomic distance--leads to a higher sensitivity to atomic position fluctuations.

For $^{133}$Cs, we observe a smooth infidelity landscape with infidelities spanning from $\sim 10^{-2}$ for low $n$ and $\Omega$ to values below $10^{-3}$ when $n$ and $\Omega$ are increased.
This trend is mostly expected from the contributions of the individual noise sources [see Figs.~\ref{fig:iswap_Sr_Cs_noise_analysis}(b) and \ref{fig:iswap_Sr_Cs_noise_analysis}(c)]:
Since we fix $\tau\Omega = {\rm const.}$, the gate duration is reduced when $\Omega$ is increased. With that, the time atoms spend in the Rydberg state
\begin{equation}
    T_{\rm ryd} = \int_0^{\tau}dt \sum_{i=1, 2}\left[ P_{\ket{r}_i}(t) + P_{\ket{r^{\prime}}_i}(t) \right] \ ,
\end{equation}
where $P_{\ket{r(r')}}(t)$ is the population of Rydberg state $\ket{r(r')}$ at time $t$, is also reduced and a decreasing impact of decay noise on the gate fidelity is expected~\cite{Pagano2022, Poole2025, Kazemi2025}.
Such a trend is clearly visible in Fig.~\ref{fig:iswap_Sr_Cs_noise_analysis}(c), where we plot the gate infidelity and the individual infidelity contributions from the atomic noise sources for fixed $n$ as a function of Rabi frequency $\Omega$.
In Fig.~\ref{fig:iswap_Sr_Cs_noise_analysis}(b), we observe that decay noise also diminishes with the principal quantum number $n$ even when $\Omega$ (and thus $\tau$) is fixed. 
The reason for this is the reduction of Rydberg state decay rates when the principal quantum number $n$ is increased, in particular, $\Gamma_{r (r')}^{\rm eff} \propto n^{-m}$, where $m \approx 3$ \cite{Saffman2010, Browaeys2016} (see appendix \ref{sec:App/FSQubitDecay} for the details).

On the other hand, interaction noise (both of the dipolar and vdW terms) impacts the system only when both atoms are simultaneously in the Rydberg states $\ket{r}$ or $\ket{r'}$, and its impact is expected to decrease when interatomic distance is enlarged, because the relative position fluctuation $\Delta R/R$ decreases.   
In our analysis, $R$ is set implicitly by $n$ and $\Omega$ through the constraint $V_{\rm dipole}/\Omega = \rm const.$ and $V_{\rm dipole} = C_3/R^3 \propto n^4/R^3$.
Therefore, increasing $n$ requires a larger $R$ and reduces the sensitivity to interaction noise, as observed in Fig.~\ref{fig:iswap_Sr_Cs_noise_analysis}(b).
In contrast, increasing $\Omega$ also requires $V_{\rm dipole}$ to increase and therefore reduces $R$ and enhances the impact of interaction noise, as shown in Fig.~\ref{fig:iswap_Sr_Cs_noise_analysis}(c).
The most robust pulses to interaction noise are consequently those at high $n$ and low $\Omega$.

Finally, we observe that Doppler noise remains mostly constant with varying $n$ [see Fig.~\ref{fig:iswap_Sr_Cs_noise_analysis}(b)]. However, it is strongly suppressed with an increasing Rabi frequency $\Omega$, as shown in Fig.~\ref{fig:iswap_Sr_Cs_noise_analysis}(c). We attribute this behavior to a direct relationship between the pulse's sensitivity to detuning fluctuations and its overall duration $\tau$, since we fixed $\tau \Omega = {\rm const.}$ in our analysis.

Overall, we observe that gate fidelities in $^{133}$Cs are dominated by atomic noise sources rather than the base vdW-inclusive optimization infidelities. Higher Rabi frequency $\Omega$ and higher principal quantum number $n$ both independently improve robustness to atomic noise, with total infidelities reaching the order of $10^{-4}$ for $n=100$ and $\Omega = 2\pi \times 50$~MHz. Importantly, even for smaller $n$ and $\Omega$ we find a large regime for which infidelities below $10^{-3}$ can be obtained.

We apply the same noise analysis for the results previously obtained from vdW-inclusive optimization for $^{88}$Sr in Fig.~\ref{fig:iswap_Sr_Cs_results_analysis}(f), but cannot obtain pulses with total infidelities below $10^{-3}$ throughout the considered $n$ and $\Omega$ parameter space.
The main reason for this is the strongly increased interaction noise compared to $^{133}$Cs, arising from two factors: (1) the slightly reduced interatomic distances required for fast, high-fidelity gate operations [cf. Fig.~\ref{fig:iswap_Sr_Cs_noise_analysis}(e)]; and (2) pulses exhibiting a longer cumulative time during which both atoms simultaneously occupy Rydberg states and interaction noise is active.
Numerically, we find that the latter is related to the stronger vdW interaction contributions in $^{88}$Sr.

Importantly, we also find that these interaction times--and therefore interaction noise--can be reduced via noise-aware optimization~\cite{Kazemi2025}.
In particular, we introduce an additional penalty on the total time spent in interacting Rydberg pair states, $T_{\rm int} = T_{rr'} + T_{rr} + T_{r'r'}$, to the optimization cost function
\begin{equation}
    C_\beta \rightarrow C_{\beta, \gamma} = C_\beta + \gamma \, T_{\rm int},
\end{equation}
where $\gamma$ is a meta-parameter defining the relative weight of $T_{\rm int}$ within the cost function, and we define
\begin{align}
    &T_{\rm rr'} = \int_0^{\tau}dt \left[ P_{\ket{rr^{\prime}}}(t) + P_{\ket{r^{\prime}r}}(t) \right] \ , \nonumber \\
    &T_{\rm rr} = \int_0^{\tau}dt P_{\ket{rr}}(t) \ ,\ \ T_{\rm r^{\prime}r^{\prime}} = \int_0^{\tau}dt P_{\ket{r^{\prime}r^{\prime}}}(t)\ ,
\end{align}
where $P_{\ket{ab}}(t)$ denotes the population of the pair state $\ket{ab}$ during the time evolution. This penalty directly suppresses the accumulation of large interaction phases during the optimization, reducing the sensitivity to positional fluctuations at the cost of higher noiseless infidelity.
We then recompute the vdW-inclusive optimizations for $^{88}$Sr using the noise-aware cost function for the same parameter space as depicted in Fig.~\ref{fig:iswap_Sr_Cs_results_analysis}(f) with $\gamma=1.0$ and apply the atomic noise calculations on the so-obtained noise-aware optimal pulses. 
The total infidelity under atomic noise is depicted in Fig.~\ref{fig:iswap_Sr_Cs_noise_analysis}(d) and is strikingly different to the result for $^{133}$Cs [Fig.~\ref{fig:iswap_Sr_Cs_noise_analysis}(a)]. 
First, the high-infidelity curve around $J_{\rm{vdW}}^{rr'}/V_{\rm dipole} \sim 1$ is retained as expected and is dominated by the infidelity from the noise-free optimization [see also peaks in Figs.~\ref{fig:iswap_Sr_Cs_noise_analysis}(e) and \ref{fig:iswap_Sr_Cs_noise_analysis}(f)].
In contrast to $^{133}$Cs, we do not find a high-fidelity region at large values of $\Omega$ and $n$. As shown in Figs.~\ref{fig:iswap_Sr_Cs_noise_analysis}(e) and \ref{fig:iswap_Sr_Cs_noise_analysis}(f), the reason for this is the large, dominating interaction noise contribution that increases with $\Omega$, as discussed above, and shows no clear trend of diminishing with $n$ (in contrast to $^{133}$Cs) due to the implications of the noise-aware optimization approach.
Importantly, however, the noise-aware optimization manages to reduce the infidelity from interaction noise clearly below $10^{-3}$ for extended regions in parameter space.

The decay noise contribution to the infidelity follows the expected trend of diminishing both with $n$ and $\Omega$ also for $^{88}$Sr. 
Doppler noise again diminishes quickly with reduced gate duration $\tau$ and therefore with increasing $\Omega$ [Fig.~\ref{fig:iswap_Sr_Cs_noise_analysis}(f)], and is rather constant as a function of $n$, however, with some more fluctuations due to the more different nature of optimized pulses throughout the full landscape.

\begin{table}[h]
    \centering
    \caption{
    Experimental parameters and noise budget obtained in the noise analysis for the
    $^{133}\rm{Cs}$ and $^{88}\rm{Sr}$ setups. The last row represents the
    infidelity of the gate when all sources of noise are simultaneously acting.
    }
    \begin{tabular*}{\columnwidth}{@{\extracolsep{\fill}} l l l l @{}}
        \toprule
        \toprule
        Parameter & & \textbf{$^{133}$Cs} & \textbf{$^{88}$Sr} \\
        \midrule
        Trap frequency & $\omega_{\rm xy}/(2\pi)$ & $100$ kHz & $200$ kHz \\
        Trap frequency & $\omega_{\rm z}/(2\pi)$ & \multicolumn{2}{c}{$40$ kHz} \\
        Laser wavevector & $k^{x}_{\rm eff}$ & \multicolumn{2}{c}{$3\times 10^{6}\ \rm{m}^{-1}$} \\
        Laser wavevector & $k^{y,z}_{\rm eff}$ & \multicolumn{2}{c}{$0$} \\
        Temperature & $T$ & \multicolumn{2}{c}{$1\ \si{\micro\kelvin}$} \\\\

        Rabi frequency & $\Omega/(2\pi)$ & $20$ MHz & $17$ MHz \\
        Rydberg level & $n$ & $80$ & $82$ \\
        Decay rate & $\Gamma_{r}/(2\pi)$ & $11.4$ kHz & $4.1$ kHz \\
        Decay rate & $\Gamma_{r'}/(2\pi)$ & $4.3$ kHz & $1.5$ kHz \\\\

        Duration & $\tau$ & $0.10\ \si{\micro\second}$ & $0.17\ \si{\micro\second}$ \\
        Distance & $R$ & $10.30\ \si{\micro\meter}$ & $8.65\ \si{\micro\meter}$ \\
        DDI & $V_{\rm dipole}/(2\pi)$ & $12.7$ MHz & $8.1$ MHz \\
        vdW interaction & $J_{\rm rr}/(2\pi)$ & $2.7$ MHz & $14.0$ MHz \\
        & $J_{\rm r'r'}/(2\pi)$ & $-0.3$ MHz & $-2.4$ MHz \\
        & $J_{\rm rr'}/(2\pi)$ & $-0.5$ MHz & $13.5$ MHz \\\\

        \multicolumn{4}{l}{Individual noise contributions $\quad 1-\mathcal{F}$} \\

        Interaction & & $4.7\times 10^{-5}$ & $2.2\times 10^{-4}$ \\
        Doppler & & $5.9\times 10^{-5}$ & $2.0\times 10^{-4}$ \\
        Decay & & $7.2\times 10^{-4}$ & $2.7\times 10^{-4}$ \\
        Laser frequency & & $2.8\times 10^{-6}$ & $2.6\times 10^{-7}$ \\
        Laser intensity & & $5.1\times 10^{-5}$ & $2.9\times 10^{-5}$ \\\\

        Gate fidelity & $\mathcal{F}$ & $99.90^{+0.015}_{-0.035}\%$
        & $99.91^{+0.042}_{-0.096}\%$ \\
        \bottomrule
        \bottomrule
    \end{tabular*}
    \label{tab:parameters_experiment}
\end{table}

Overall, the noise-aware optimization allows us to find pulses with total fidelities below $10^{-3}$ under atomic noise, also for $^{88}$Sr. 
However, compared to $^{133}$Cs, the corresponding parameter regime is much smaller, mainly due to the strongly increased contribution from interaction noise.
At fixed $n=82$, total infidelities below $10^{-3}$ are achieved for the range $\Omega/(2\pi) \in [15,\ 18]$ MHz while at fixed $\Omega/(2\pi) = 17$ MHz the corresponding window in principal quantum number spans only $n \in [82,\ 86]$. 

Based on this analysis, we individually select a pulse with total infidelity $1-\mathcal{F} \lesssim 10^{-3}$ for the two considered species. While this choice is rather limited for $^{88}$Sr, for $^{133}$Cs we additionally assume experimentally feasible values of $\Omega = 2\pi \times 20$~MHz and $n=80$.
The selected pulses are shown in Fig.~\ref{fig:fig1_schemes}(b), and full parameters for those selected pulses are listed in Table~\ref{tab:parameters_experiment}.
Additionally, in Fig.~\ref{fig:fig1_schemes}(b) we plot the population dynamics of the states $\ket{0}, \ket{1}, \ket{r}, \ket{r'}$ on the first atom during the pulse sequence after initializing the atoms in the state $\ket{0} \otimes \ket{1}$. We observe that the entire population at the end of the pulse sequence has exchanged to state $\ket{1}$, as expected for a successful implementation of the iSWAP gate.
We consider only those two pulses for the subsequent analysis.

\begin{figure}[t]
    \centering
    \includegraphics[width=\columnwidth]{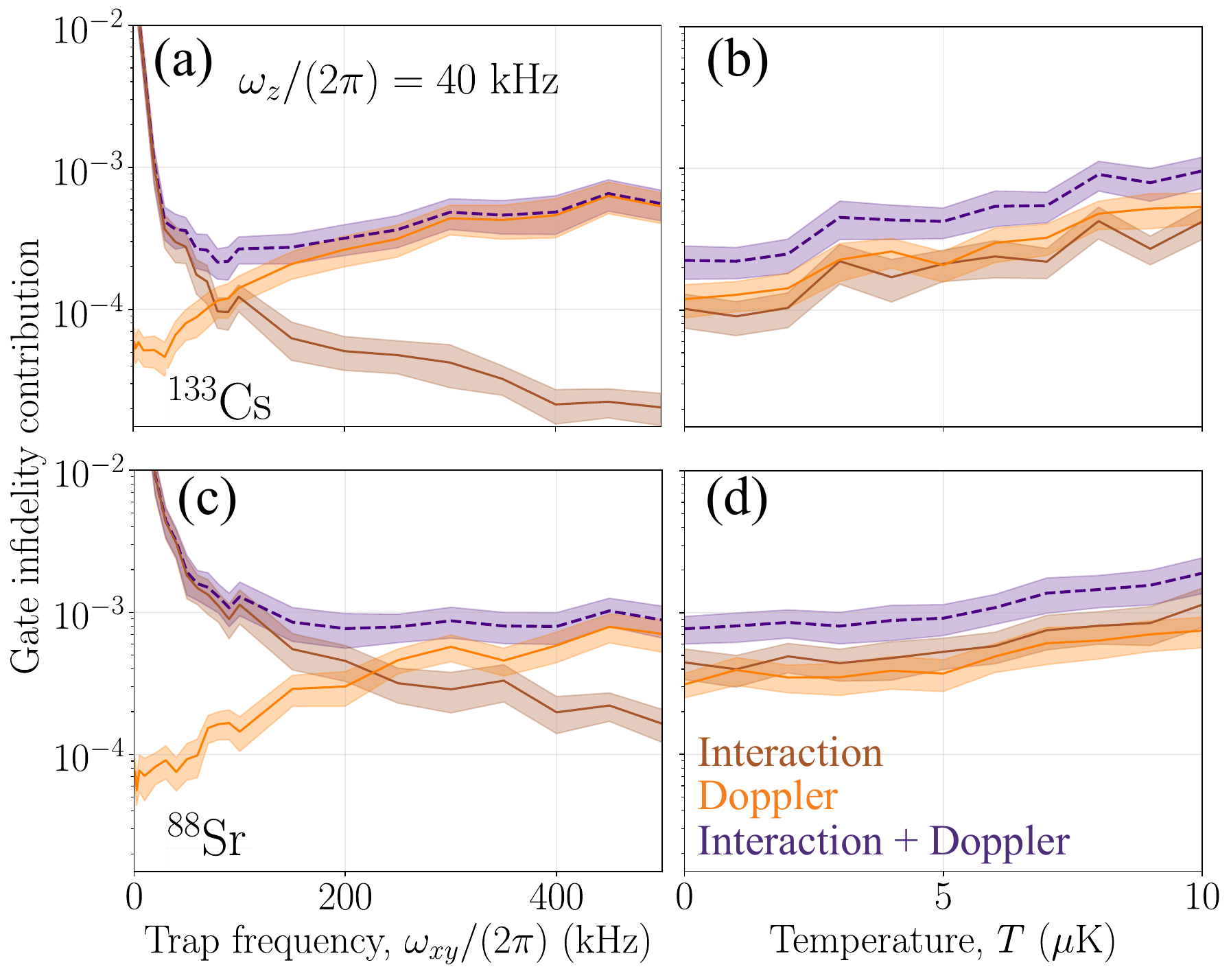}
    \caption{Trap frequency and temperature dependence of vdW-inclusive optimized pulses for $^{133}$Cs [panels (a) and (b)] and $^{88}$Sr [panels (c) and (d)]. 
    Individual contributions of Doppler and interaction noise on the overall gate infidelity are plotted as a function of trap frequency $\omega_{xy}$ [panels (a) and (c)] for fixed $T=1 \, \si{\micro\kelvin}$, and as a function of temperature $T$ of the atoms for fixed $\omega_{xy}=2\pi \times 100$~kHz [panel (b)] and $\omega_{xy}=2\pi \times 200$~kHz [panel (d)].
    The combined contribution of Doppler and interaction noise is shown by the dashed line.
    In all plots, we use $\omega_z = 2\pi \times 40$~kHz and the species-specific optimal pulses and parameters selected above [cf. Fig.~\ref{fig:fig1_schemes}(b) and Table~\ref{tab:parameters_experiment}].
    }
    \label{fig:iswap_Sr_Cs_trap_temp_analysis}
\end{figure}
 
Having used fixed parameters for the trap frequencies $\omega_{xy}$ and the temperature $T$ of the atoms for the analysis in Fig.~\ref{fig:iswap_Sr_Cs_noise_analysis}, we now analyze the influence of these parameters on the infidelity contributions of interaction and Doppler noise for the species-specific pulses selected above.
Our results are shown in Fig.~\ref{fig:iswap_Sr_Cs_trap_temp_analysis}. 
We clearly observe in Figs.~\ref{fig:iswap_Sr_Cs_trap_temp_analysis}(a) and \ref{fig:iswap_Sr_Cs_trap_temp_analysis}(c) that interaction noise is dominant for shallow traps with small $\omega_{xy}$ where atom position fluctuations are large, while Doppler noise becomes dominant in tighter traps with larger $\omega_{xy}$ because interaction noise is reduced due to tighter confinement, at the expense of increasing the average velocity of atoms and thereby the Doppler noise. 
The combined contribution of interaction and Doppler noise is shown by the dashed lines and obeys a minimum because of the trade-off between the two noise types.
The optimal value depends on the species because of the different mass, but also because of different pulse profiles and the resulting different relative strengths between the interaction and Doppler noise contributions. For the two candidate species and respective optimal pulses, the optimal values are $\omega_{xy} \approx 2\pi \times 100$ kHz for $^{133}$Cs, and $\omega_{xy} \approx 2\pi \times 200$ kHz for $^{88}$Sr [cf. Figs.~\ref{fig:iswap_Sr_Cs_trap_temp_analysis}(a) and \ref{fig:iswap_Sr_Cs_trap_temp_analysis}(c)]
On the contrary, both interaction and Doppler noise contributions increase when the temperature $T$ of the atoms is increased [see Figs.~\ref{fig:iswap_Sr_Cs_trap_temp_analysis}(b) and \ref{fig:iswap_Sr_Cs_trap_temp_analysis}(d)]. For $^{133}$Cs the combined infidelity contribution remains below $10^{-3}$ for $T \lesssim 10 \, \si{\micro\kelvin}$, whereas for $^{88}$Sr the combined infidelity exceeds $10^{-3}$ already at $T\sim 5 \, \si{\micro\kelvin}$, mainly due to the increased interaction noise effects.

\subsection{Laser noise}
\label{sec:laser_noise}

\begin{figure*}[t]
    \centering
    \includegraphics[width=\textwidth]{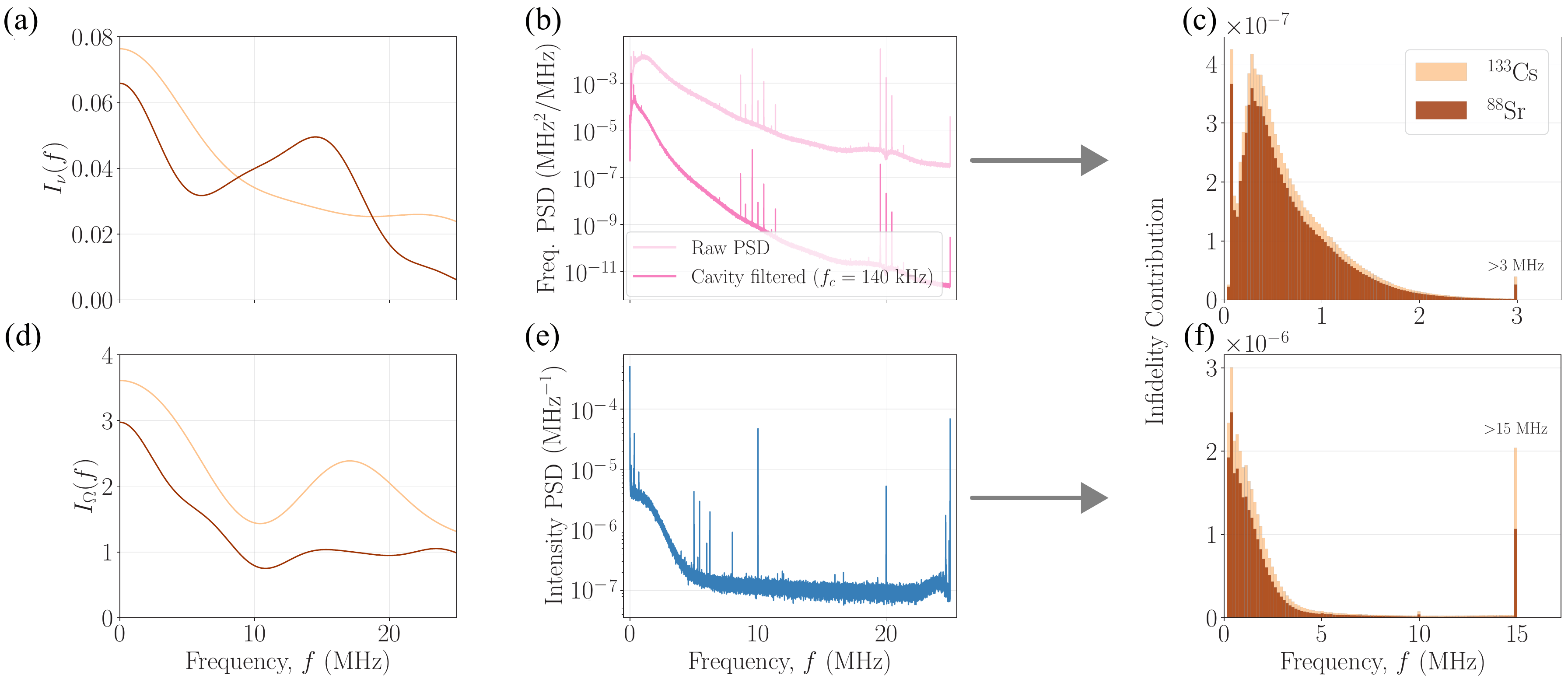}
    \caption{
    FRT-based analysis of previously selected pulses for $^{133}$Cs and $^{88}$Sr [cf. Fig.~\ref{fig:fig1_schemes}(b)].
    FRT analysis of laser frequency noise showing (a) the pulses' RF $I_\nu(f)$, (b) the corresponding PSD, and (c) the resulting infidelity contributions as a function of the noise frequency $f$, obtained from multiplying the RF by the PSD. The total contribution of the laser frequency noise on the infidelity (within the FRT approximation) is given by integration over $f$.
    In panel (b) we show the raw laser frequency PSD (light pink) and the same one filtered through a cavity with $140$ kHz linewidth (pink). To compute the infidelity contribution in panel (c), we consider the cavity-filtered PSD.
    (d)-(f) Same analysis for laser intensity noise.
    Different colors in panels (a), (c), (d), and (f) correspond to the different pulse protocols chosen for the two atomic species.
    }
    \label{fig:pulses_analysis_response_function_vdw}
\end{figure*}

Having characterized the sensitivity of the optimized pulses to atomic noise sources, we now turn to laser noise. Specifically, we investigate the effect of temporal fluctuations in the intensity and frequency of the driving fields, characterized through their PSDs, on the previously selected optimal pulses [see Fig.~\ref{fig:fig1_schemes}(b) and Sec.~\ref{sec:atomic_noise}]. 
The PSD quantifies the distribution of noise power across frequencies, and thereby encodes the temporal correlation structure of the fluctuations, determining how strongly a given pulse, with its particular spectral response, couples to the laser noise~\cite{Endres2025Benchmark}.
As a realistic example, we assume previously measured PSDs for the laser frequency and intensity noise from Ref.~\cite{Endres2025Benchmark} [see also Figs.~\ref{fig:pulses_analysis_response_function_vdw}(b) and \ref{fig:pulses_analysis_response_function_vdw}(e)].
For simplicity, we assume identical intensity and frequency PSDs for all drive terms, and for the frequency PSD, we assume that the laser is additionally cavity-filtered~\cite{Endres2025Benchmark} with a cavity linewidth of $140$ kHz [see Eq.~\eqref{eq:cavity_filtering} for details].
The noise effect from a PSD on the gate fidelities can be evaluated in a Monte Carlo sampling-based approach~\cite{deLeseleuc2018, Jiang2022}, which modifies the time-dependent Rabi frequencies $\Omega^{ab}(t) \rightarrow \tilde\Omega^{ab}(t)$, and frequencies $\nu^{ab}(t) \rightarrow \tilde\nu^{ab}(t)$ for each shot (see Appendix~\ref{App:phase} for details). Given our definition of the driving terms in Eq.~\eqref{eq:drive_hamiltonian} in terms of phases $\phi$ instead of frequencies $\nu$, and the relationship between frequency and phase PSDs, $S_{\phi}(f) = S_{\nu}(f)/(f^2)$, we model frequency noise as fluctuations in the $\phi^{ab}(t) \rightarrow \tilde\phi^{ab}(t)$ based on the phase PSD, $S_{\phi}(f)$.

The so-obtained contributions of laser frequency and intensity noise on the gate infidelity for the previously selected pulses for $^{133}$Cs and $^{88}$Sr are shown in Fig.~\ref{fig:fig1_schemes}(d).
Both pulses exhibit rather high robustness to laser frequency noise with infidelity contributions around or below $10^{-6}$. 
Contributions from laser intensity noise are larger, reaching values on the order of $10^{-5}$.
Overall, laser-noise-induced infidelities remain at least an order of magnitude smaller than the corresponding atomic-noise contributions, indicating that laser noise is not the dominant limitation for either protocol.

To understand the origin of the differences between laser frequency and intensity noise, and to gain deeper insight into how each pulse couples to noise at specific frequencies, we employ the recently developed \textit{fidelity response theory} (FRT)~\cite{Endres2025Benchmark}. FRT constructs a pulse response function (RF) that characterizes the sensitivity of a given pulse shape to a specific type of noise at a given frequency. Combined with the corresponding PSD, the RF enables the determination (to first order) of how much frequency or intensity noise at each frequency spectral component contributes to the fidelity loss. More precisely, within the FRT framework, the infidelity for a given pulse due to a laser noise channel $\alpha$ (where $\alpha$ denotes intensity or frequency noise) is approximated as
\begin{equation}
1 - \mathcal{F}_{\alpha} = \int_0^{\infty} df \, S_\alpha(f) \, I_\alpha(f) \ ,
\label{eq:FRT_fidelity}
\end{equation}
where $S_\alpha(f)$ is the PSD at noise frequency $f$ and $I_\alpha(f)$ is the corresponding RF, evaluated as~\cite{Endres2025Benchmark}
\begin{align}
I_\alpha(f) = \int_0^T \int_0^T &dt \, d\tau \, \cos\big(2\pi f (t-\tau)\big) \nonumber \\
&\times \langle \hat{O}_\alpha^H(t), \hat{O}_\alpha^H(\tau)\rangle_c \, .
\end{align} 
Here, $\hat{O}_\alpha^H(t)$ denotes the noise operator of channel $\alpha$ in the Heisenberg picture and $\langle \hat A , \hat B \rangle_c$ represents the connected correlation function between operators $\hat A$ and $\hat B$. Detailed derivations of the RFs are provided in Appendix~\ref{App:FRT}. 

Figure~\ref{fig:pulses_analysis_response_function_vdw} shows the FRT analysis for the two selected pulse protocols. In particular, Fig.~\ref{fig:pulses_analysis_response_function_vdw}(a) displays the frequency noise RF $I_{\nu}(f)$ for both protocols (distinguished by color). 
The two pulses exhibit qualitatively distinct behaviors. The $^{88}$Sr pulse, which features a longer gate duration, exhibits a localized sensitivity peak near $15$~MHz. In contrast, the $^{133}$Cs pulse displays pronounced sensitivities at both low ($f \lesssim 10$~MHz) and high ($f \gtrsim 20$~MHz) frequencies. Importantly, however, both pulses feature significantly small RF magnitudes overall $(I_{\nu}\lesssim 0.08)$, demonstrating the intrinsic robustness of these iSWAP gate protocols against laser frequency fluctuations.

Figure~\ref{fig:pulses_analysis_response_function_vdw}(b) depicts the corresponding raw and cavity-filtered frequency PSDs, both of which decay strongly with frequency $f$. The frequency-resolved infidelity contributions are obtained by multiplying the RF by the PSD. The result, presented in Fig.~\ref{fig:pulses_analysis_response_function_vdw}(c), demonstrates that infidelity contributions from (quasi)static, shot-to-shot frequency noise ($f\rightarrow 0$) are small for both pulses. Instead, the contribution peaks on the order of $10^{-7}$ at small but nonzero frequencies, driven by the underlying profile of the PSD. 
Due to the sharp decay of the PSD, frequency contributions above $f\gtrsim 2$~MHz become negligible. Consequently, even though the pulses feature distinct RF spectral profiles—with $^{88}$Sr showing an isolated sensitivity peak at $f \sim 15$~MHz in the RF [cf. Fig.~\ref{fig:pulses_analysis_response_function_vdw}(a)], the low magnitude of the cavity-filtered PSD ($<10^{-5}$) in this high-frequency regime suppresses this feature. This leads to qualitatively similar distributions of infidelity contributions across the $10$-$20$ MHz frequency range, with $^{133}$Cs exhibiting a marginally larger overall sensitivity to this noise channel.

An identical analysis for laser intensity noise is presented in Figs.~\ref{fig:pulses_analysis_response_function_vdw}(d)-\ref{fig:pulses_analysis_response_function_vdw}(f). Again, the intensity RFs $I_{\Omega}(f)$ differ qualitatively between the two pulses. Similar to the frequency RFs, $I_{\Omega}(f)$ peaks at low $f$, indicating that the pulses lack intrinsic robustness against (quasi)static intensity fluctuations. The intensity PSD [Fig.~\ref{fig:pulses_analysis_response_function_vdw}(e)] also diminishes at high frequencies, albeit more slowly than the frequency PSD. Together, these features dictate that laser intensity noise contributions to the infidelity are primarily accumulated at frequencies $f \lesssim 5$~MHz, dominated by quasistatic contributions as $f \rightarrow 0$. 

While integrated infidelity contributions for $f>15$~MHz are on the order of $10^{-6}$, they remain negligible compared to the integrated contributions from the low-frequency regime.
Furthermore, under the considered PSD no single frequency in this high-frequency regime shows a particularly strong contribution to the infidelity (not shown).

Overall, the $^{88}$Sr pulse shows slightly lower infidelity contributions than the $^{133}$Cs pulse, with the sum over all frequencies being in qualitative agreement with the ones obtained from non-perturbative MC simulations reported in Fig.~\ref{fig:fig1_schemes}(d) and Table~\ref{tab:parameters_experiment}.

In summary, this section has analyzed the fidelity of the proposed iSWAP gate protocols for both $^{133}\rm{Cs}$ and $^{88}\rm{Sr}$ across a wide range of experimental parameters, isolating the individual impacts of atomic and laser noise sources. For both species, we have identified optimized pulse protocols that achieve iSWAP gate fidelities exceeding $99.9\%$ under realistic noise conditions. The resulting pulse profiles and their comprehensive noise budgets are summarized in Figs.~\ref{fig:fig1_schemes}(b) and \ref{fig:fig1_schemes}(d), with exact parameters compiled in Table~\ref{tab:parameters_experiment}.

\section{Conclusions}\label{sec:conclusion}

In summary, this work demonstrates a pathway toward high-fidelity, hardware-native iSWAP entangling gates for Rydberg atoms using tailored control protocols. By concurrently implementing state transfer and interaction within a single time-dependent control pulse and leveraging quantum optimal control methods, we achieved high-fidelity protocols for the iSWAP gate for different driving schemes and both Rabi-modulated and phase-modulated pulses.
Our approach considers global controls and smooth pulse shapes to minimize experimental demands.
Even after accounting for vdW interactions, the optimized pulses maintain high fidelity and fast gate performance. Using the prototypical atomic species $^{133}{\rm Cs}$ and $^{88}{\rm Sr}$ as examples, we find that the ratio between the DDI and the vdW interaction strongly affects the physical realization of the iSWAP gate.

We performed extensive noise modeling to analyze the gate fidelities under experimentally realistic conditions. In particular, we considered noise from the motion of atoms, noise due to decay of the Rydberg states, and time-dependent laser frequency and intensity noise modeled with PSDs.
We conducted a detailed analysis of the influence of experimental parameters on the infidelity contributions of these individual noise sources for optimized pulse protocols and demonstrate that both $^{133}$Cs and $^{88}$Sr can reach iSWAP gate fidelities of $99.9\%$ under ``optimal'' parameters realistic for near-term experiments.
Furthermore, our results indicate that $^{133}\rm{Cs}$ offers a considerably larger window of experimental parameters over which this fidelity threshold is maintained, whereas the corresponding parameter space for high-fidelity gate operation in $^{88}\rm{Sr}$ is substantially more restricted.

We have also utilized the FRT framework to analyze the specific response of different pulses to laser frequency and intensity noise. 
FRT provides valuable insight into the laser noise susceptibility of different optimal pulse protocols and could be utilized to select pulses that perform particularly well for a given set of PSDs in future work.
Conversely, the same framework could be used to adjust laser PSDs to a given pulse response function to further reduce noise in a tight co-design approach.

Looking toward experimental realization, we briefly note several other effects that could influence the gate fidelity. 
While we have thoroughly considered the second-order vdW interactions in this work, third-order contributions from the dipole-dipole interaction, scaling as $1/R^9$, can become non-negligible and induce additional phase shifts or state mixing, particularly at short distances~\cite{Cohen2021}. 
However, for the optimal parameter range identified in this article $(R \approx 8.65-10.3\ \si{\micro\meter})$, we do not expect significant contributions from such terms. 
Additionally, because the Rydberg level spacing scales as $n^{-3}$, operating at high principal quantum numbers $n$ with strong Rabi driving $\Omega$ could give rise to off resonant excitations into adjacent Rydberg states. 
Importantly, our results demonstrate high-fidelity gate protocols already for $n \gtrsim 60$ [cf.~Figs.~\ref{fig:iswap_Sr_Cs_noise_analysis}(a) and \ref{fig:iswap_Sr_Cs_noise_analysis}(d)]. 
Furthermore, these multi-level leakage channels can be mitigated by appropriate choices of the Rydberg states or by converting leakage into detectable erasure errors \cite{Ma2023_erasure, Scholl2023}.
Dipole-dipole interactions can also induce state-flipping leakage between magnetic sublevels within the $\ket{r}$ and $\ket{r'}$ manifolds. These transitions can be straightforwardly rendered off resonant by applying an appropriate magnetic field to lift the Zeeman degeneracy.
Finally, electric field fluctuations can impose operational constraints. Although our scheme does not require a static electric field to bring the exchange interaction on resonance ($\Delta_F=0$ through the choice of $\ket{r}$ and $\ket{r'}$), residual stray fields could still reduce gate fidelity: because $\ket{r}$ and $\ket{r'}$ generically have different dc polarizabilities, field noise can perturb the exchange coupling out of resonance. 
In practice, such electric stray fields can be suppressed, for example, through the use of electric shielding.

Our findings highlight the potential to extend the neutral-atom gate set beyond the usually considered diagonal entangling gates by using and controlling a second Rydberg state.
Extending our approach to include atomic and laser noise directly in the optimization procedure could further improve gate fidelities and enhance robustness for specific experimental neutral-atom QPU platforms.
To this end, noise-aware optimization cost functions~\cite{Kazemi2025} and the FRT approach to laser noise~\cite{Endres2025Benchmark} could be very useful to reduce noise sampling costs throughout the optimization process.
Another interesting direction would be the design of specifically tailored, noise-robust iSWAP gate implementations using previously established techniques~\cite{janduraTimeOptimalTwoThreeQubit2022, Fromonteil2023}.
Considering the entire exchange gate family, $U_{\rm XY}(\theta)$, finding analytically parametrized pulses for the angle $\theta$ would simplify the implementation of a continuous set of gates by reducing experimental calibration efforts~\cite{Evered2023}.

\begin{acknowledgments}
We acknowledge insightful discussions with Johannes Zeiher and Andrea Alberti on experimental realization with strontium qubits, with Sylvain de L{\'e}s{\'e}leuc on laser phase noise, with Sebastian Weber on fidelity response theory, and with both Johannes Mögerle and Sebastian Weber on atomic physics calculations within pairinteraction.
We especially thank Manuel Endres for providing the data for the laser phase and intensity PSDs.
This project has received funding from the
European Union (Horizon-MSCA-Doctoral Networks)
through the project QLUSTER (HORIZON-MSCA--2021-DN-01--GA101072964). This research was funded by the German Federal Ministry of Research, Technology and Space (BMFTR) within the project MUNIQC-ATOMS (Project No. 13N16080).
This study was supported by the Austrian Research Promotion Agency (FFG Project No. FO999924030, FFG Basisprogramm).
This publication has received funding under Horizon Europe programme HORIZON-CL4-2022-QUANTUM-02-SGA via the Project 101113690 (PASQuanS2.1).
\end{acknowledgments}

\bibliography{bibliography_exchange_gate}

\appendix

\section{Parametrized exchange gate}
\label{app:exchange_gate}

The exchange interaction between the two Rydberg states $\ket{r}$ and $\ket{r'}$ allows the implementation of parametrized \textit{exchange} gates 
$U_{\rm XY}(\theta)$ in the qubit manifold, where $\theta$ denotes the gate angle. 
In the main text, we focused our discussion on the iSWAP gate, which is realized by $U_{\rm XY}(\theta=\pi)$.
The same techniques used there can be applied for arbitrary angles $\theta$.

\begin{figure}[t]
    \centering
    \includegraphics[width=1\columnwidth]{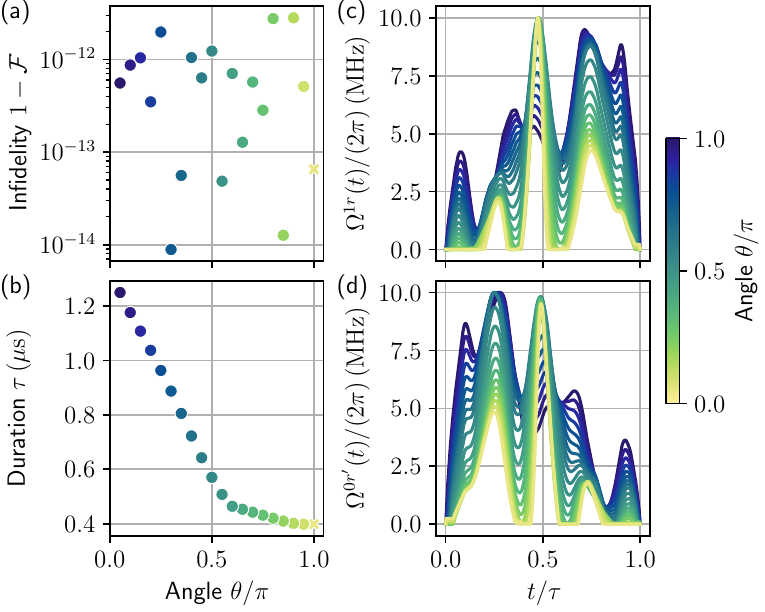}
    \caption{
    Parametrized exchange gate optimization. Optimal pulses for different angles $\theta$ are obtained by sequential optimization starting from the iSWAP gate result at $\theta=\pi$ and fixing the maximal Rabi frequency $\Omega = 2\pi \times 10 \ {\rm MHz}$.
    (a) Infidelity of $U_{\rm XY}(\theta)$, and (b) gate duration $\tau$ plotted vs the gate angle $\theta$.
    Optimized pulse drives, (c) $\Omega^{1r}(t)$, (d) $\Omega^{0r'}(t)$, plotted vs relative time $t/\tau$ to illustrate the continuous pulse deformation with the angle $\theta$.
    }
    \label{fig:exchange_gate_angles}
\end{figure}

Here, we adopt a slightly different strategy: We use an optimal pulse for the iSWAP gate with $\theta=\pi$, and then use this result as the initial guess for the optimization at a slightly smaller angle.
We continue this strategy to obtain a sequence of pulse protocols for $\theta \in \left(0, \pi\right]$.
Within this section, we focus on vdW-free, driving scheme A with Rabi-modulated pulses [see Fig.~\ref{fig:pulse_landscapes_all}(c)] and choose a high-fidelity pulse near the speed limit as our initial optimal protocol for the iSWAP gate.
Figure~\ref{fig:exchange_gate_angles}(a) shows the resulting infidelities, all of which remain below $10^{-11}$. Fixing the maximal Rabi frequency $\Omega$ leads to increased pulse durations for angles $\theta < \pi$ for the given Rabi-modulated pulse, as shown in Fig.~\ref{fig:exchange_gate_angles}(b). 
The main advantage of this approach is evident from the pulse protocols for the different angles $\theta$, depicted in Figs.~\ref{fig:exchange_gate_angles}(c) and \ref{fig:exchange_gate_angles}(d), which show a continuous deformation when the angle is changed.
This suggests that high-fidelity pulses across the continuous range $\theta \in \left(0, \pi\right]$ could be obtained through interpolation techniques.
Furthermore, experimental calibration for the different angles could be simplified, and noise characteristics are expected to be qualitatively similar.

The results shown here demonstrate that even this simple protocol allows to generate an entire family of high-fidelity exchange gates $U_{\rm XY}(\theta)$. 
The observed increase in the gate duration $\tau$ with decreasing $\theta$ is likely an artifact of the sequential optimization strategy employed here, since the iSWAP gate with $\theta=\pi$ is the most entangling gate in the exchange gate family $U_{\rm XY}(\theta)$. 
Therefore, we expect that faster pulses for $\theta < \pi$ could be identified by performing independent optimizations from randomly initialized ansatzes. However, this would most likely lead to qualitatively different pulse shapes.
More elaborate approaches that identify families of similar pulses and allow for their interpolation~\cite{deKeijzer2024} could be investigated in future work to overcome some of these limitations.

\section{State Transfer Protocol}
\label{app:ST_protocol}

\begin{figure}[t!]
    \centering
    \includegraphics[width=\columnwidth]{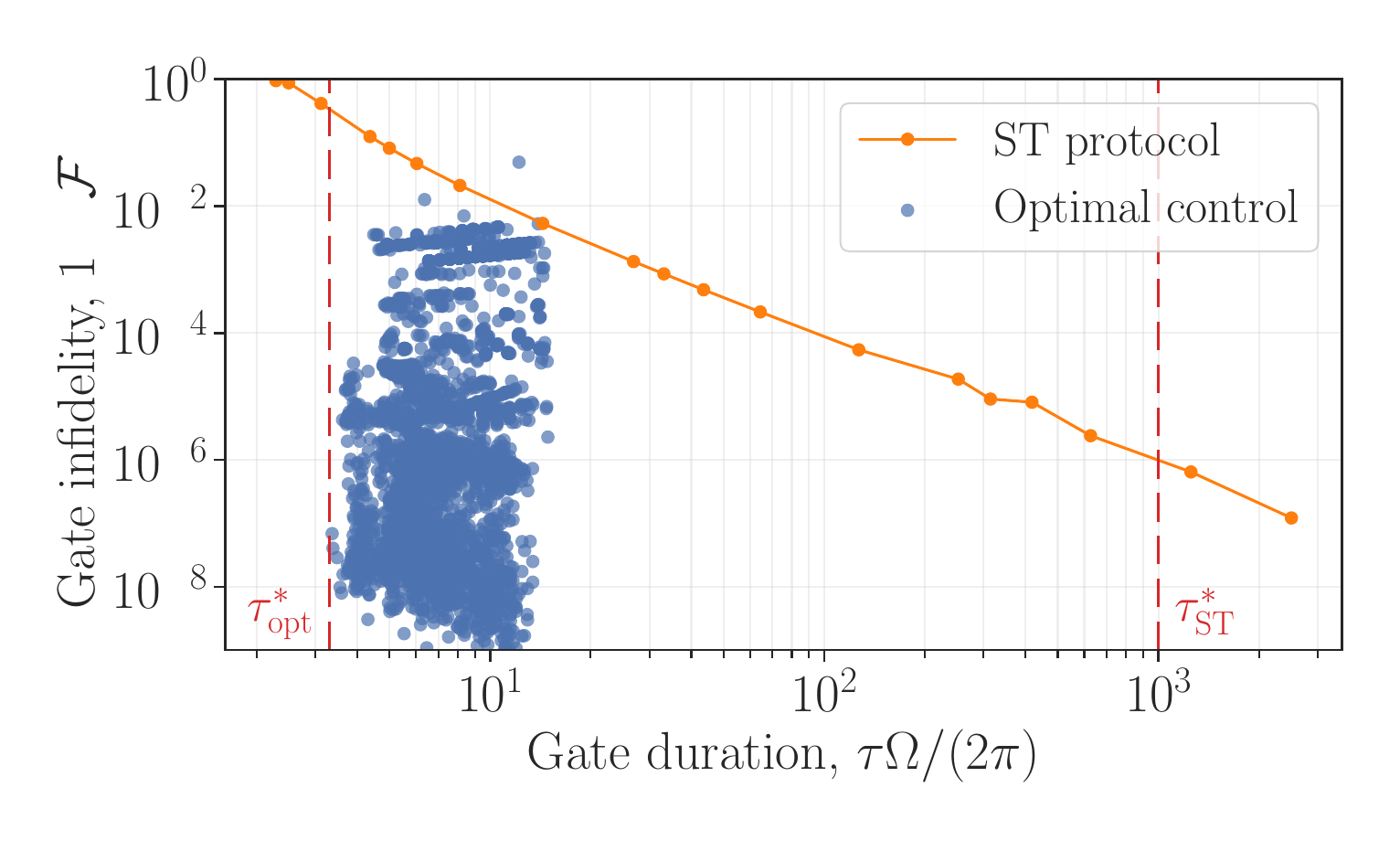}
    \caption{Comparison of the gate infidelity as a function of dimensionless gate duration between Rabi-modulated, optimally controlled pulses and the ST protocol (introduced in Sec.~\ref{sec:setup}) for an iSWAP gate, assuming an ideal, vdW-free setup. Optimal control allows the gate to reach the speed-limit threshold ($1-\mathcal{F} < 10^{-6}$) at a gate duration roughly 300 times shorter than the standard ST protocol.}
    \label{fig:opt_vs_ST}
\end{figure}

The ST protocol implements exchangelike gates, such as the iSWAP, via a three-step sequence: a Ramsey-like pulse, a period of pure DDI, and a final Ramsey-like pulse. However, this protocol relies on Rydberg interactions remaining negligible during the driving pulses. One way to satisfy this condition is by operating in a regime where the Rabi frequency of the driving lasers greatly exceeds the DDI strength, $\Omega \gg V_{\rm dipole}$. 

For the iSWAP gate realized via scheme A, we illustrate the performance of this protocol in Fig.~\ref{fig:opt_vs_ST} at a fixed, experimentally feasible Rabi frequency of $\Omega/(2\pi) = 10$~MHz. We plot the gate infidelity across different ratios of $\Omega/V_{\rm dipole}$, which directly map to different total gate durations $\tau$ since the sequence requires $\tau = \tau_{\pi\text{-pulse}} + \tau_{\rm dipole} + \tau_{\pi\text{-pulse}}$. Given that $\tau_{\pi\text{-pulse}}\Omega = \pi$ and $\tau_{\rm dipole}V_{\rm dipole} = \pi/2$, the dimensionless gate duration scales as
\begin{equation}
\frac{\tau\Omega}{2\pi} = 1 + \frac{1}{4} \frac{\Omega}{V_{\rm dipole}} \,.
\end{equation}
As the ratio $\Omega/V_{\rm dipole}$ decreases, the total gate duration becomes shorter at the expense of significantly higher infidelities. Defining the quantum speed limit as the shortest gate duration achieving an infidelity $1-\mathcal{F} < 10^{-6}$, the ST protocol reaches this threshold at $\tau^*_{\rm ST}\Omega/(2\pi) \approx 10^3$. In comparison, the optimal control results for Rabi-modulated pulses under scheme A show that the optimized speed limit $\tau^*_{\rm opt}$ occurs at durations approximately 300 times shorter, as previously highlighted in Fig.~\ref{fig:pulse_landscapes_all}(c). 

\section{Hardware platforms}
\label{app:hardware_platforms}

\begin{figure}[t]
    \centering
    \includegraphics[width = \columnwidth]{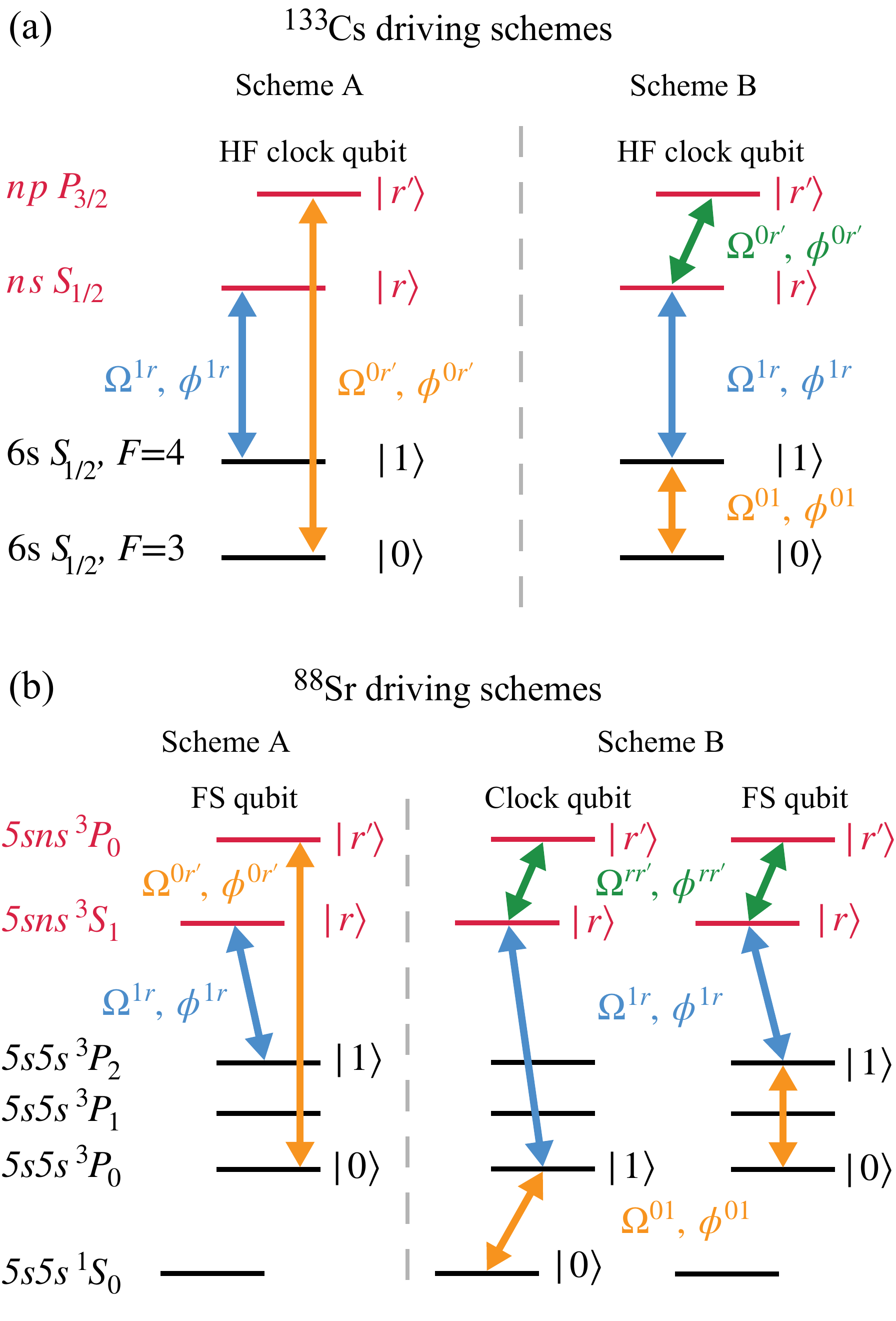}
    \caption{
    Atomic scheme proposals for typical (a) $^{133}$Cs and (b) $^{88}$Sr setups.  For $^{133}$Cs, by encoding the qubit in the hyperfine structure of the ground state (HFQ), either scheme, A or B, can be implemented. For $^{88}$Sr, we propose two different qubit encodings: the FSQ and CQ encodings. Scheme A can be implemented solely with the FSQ encoding, while scheme B can be implemented with either the FSQ or CQ encodings.
    }
    \label{fig:88Sr_133Cs_schemes}
\end{figure}

In this appendix, we outline representative implementations of driving schemes A and B for the atomic species and qubit encodings considered in the main text (see Fig.~\ref{fig:88Sr_133Cs_schemes}). Specifically, we consider the HFQ encoding of $^{133}$Cs, a well-established neutral-atom platform~\cite{Manetsch2024, anand2024dual, graham2022multi, graham2023midcircuit}, and the CQ and FSQ encodings of $^{88}$Sr~\cite{Scholl2023, Endres2025Benchmark, Tao2025}. The CQ encoding employs the ultranarrow optical clock transition~\cite{PhysRevLett.91.173005, ludlow2008sr, 5422497, PhysRevLett.103.063001, PhysRevLett.106.210801, PhysRevA.84.052716, PhysRevLett.124.203201}, while the FSQ encoding utilizes long-lived fine-structure states coupled by a two-photon Raman transition~\cite{meinert2021quantum, PhysRevLett.132.150605, PhysRevLett.132.150606}. 

For $^{133}$Cs, illustrated in Fig.~\ref{fig:88Sr_133Cs_schemes}(a), both schemes A and B can be realized within the HFQ encoding with Rydberg states $\ket{r} = \ket{ns(S_{1/2})_{m_j = +1/2}}$ and $\ket{r'} = \ket{np(P_{3/2})_{m_j = +3/2}}$. Scheme A requires two two-photon drives coupling $\ket{1}$ to $\ket{r}$ and $\ket{0}$ to $\ket{r'}$, respectively. In scheme B, the $\ket{0}\leftrightarrow\ket{r'}$ drive is replaced by a direct microwave coupling between the HFQ states $\ket{0}\leftrightarrow\ket{1}$, while the $\ket{1}\leftrightarrow\ket{r}$ transition remains two-photon driven and the Rydberg states $\ket{r}$ and $\ket{r'}$ are coupled by a microwave field.

For $^{88}$Sr, illustrated in Fig.~\ref{fig:88Sr_133Cs_schemes}(b), scheme A is considered only for the FSQ encoding, where the $\ket{0}\leftrightarrow\ket{r'}$ transition is driven via a two-photon process and the $\ket{1}\leftrightarrow\ket{r}$ transition is driven directly. Scheme B can be implemented with either the FSQ or CQ encoding. In the FSQ encoding, the qubit states are coupled via a two-photon Raman transition, whereas in the CQ encoding they are connected by a direct optical drive. In both cases, the $\ket{1}\leftrightarrow\ket{r}$ transition is driven directly and the Rydberg states are coupled by a microwave field. The considered Rydberg states are $\ket{r} = \ket{5sns(^3S_1)_{m_j = +1}}$ and $\ket{r'} = \ket{5snp(^3P_0)_{m_j = 0}}$.

The implementations described above serve as representative hardware realizations used to estimate the performance of the protocol. Alternative implementations and combinations of qubit encodings and driving schemes may also be feasible depending on the available control fields and experimental constraints. Unless stated otherwise, all results presented in the main text are obtained assuming driving scheme A with phase modulation at a fixed Rabi frequency $\Omega$.

\section{Atomic parameters for \texorpdfstring{$^{88}$Sr}{88Sr}}
\label{sec:App/FSQubitDecay}

The inverse radiative lifetime of an atomic state 
$\Gamma_i$, at finite temperature of the environment $T>0$ is given by
\begin{equation} \label{lifetime}
\Gamma_{i} = \sum_{j: E_j < E_i} \gamma_{ij} [1+n(\omega_{ij},T)] 
+
\sum_{j: E_j > E_i} \gamma_{ij} n(\omega_{ij},T),
\end{equation}
where
\begin{equation}  
\gamma_{ij}=\sum_{j} \frac{4}{3} \frac{|\langle i| \hat{\vec{D}} | j \rangle|^2}{\hbar c^3} \omega_{ij}^3,
\end{equation}
and $n(\omega_{ij},T)$ describes the influence of BBR 
\begin{equation}
n(\omega_{ij},T) = \big[\exp(\tau_R \omega_{ij}) - 1\big]^{-1}.
\end{equation}
Here, $\omega_{ij}=|E_i-E_j|/\hbar$ and  $\tau_R=\hbar/k_B T$ is so-called photon reservoir correlation time \cite{carmichael2013statistical}, the temperature entering here refers to the photon reservoir, not to the atomic subsystem.

The first sum on the right-hand side \eqref{lifetime} runs over all possible atomic states $j$ with lower energies $E(j)<E(i)$ and describes spontaneous and induced decay processes. The second sum over the higher-lying states $E(j)>E(i)$ describes the BBR-induced excitation processes, relevant for the closely spaced Rydberg states.  The variety of dipole-coupled states is determined by the electric-dipole selection rules via dipole matrix elements $\langle i| \hat{\vec{D}} | j \rangle$, whose evaluation we describe in detail below.

Here, we report about alkaline-earth atom  $^{88}$Sr. This isotope has no nuclear spin, and therefore exhibits no hyperfine structure. On the other hand, the presence of two valence electrons requires the coupling of spin and orbital angular momenta of individual electrons. This is handled using the non-relativistic $LS$-coupling scheme, 
\begin{subequations} \label{Coupling_Scheme}
\begin{align}
\vec{J} &= \vec{L} + \vec{S},  
\\
\vec{L} &= \vec{l}_1 + \vec{l}_2, 
\\
\vec{S} &= \vec{s}_1 + \vec{s}_2, 
\end{align}
\end{subequations}
where $\{ J, L, S\}$ are the total atomic quantum numbers. Atomic terms are labeled as $^{2S+1}L_J$, while $l_{1,2}$ and $s_{1,2}$ denote the quantum numbers of individual valence electrons.

Let $\hat{\vec{D}}=\hat{\vec{d}}_1+\hat{\vec{d}}_2$ be an operator of the electric dipole moment of both electrons. Assuming only one electron is excited by the incident field and neglecting two-electron excitations, we take $\hat{\vec{D}}=\hat{\vec{d}}_2$. We now evaluate the dipole moment matrix element $\langle i| \hat{\vec{D}} | j \rangle$ between a pair of states $|i\rangle = |J' M_{J'} \rangle$ and $|j\rangle =|J M_J \rangle$, where only one electron is excited (here the primed and unprimed denotations do not refer to $|r'\rangle$ and $|r\rangle$ from the main text). 
Following the coupling scheme \eqref{Coupling_Scheme}, we obtain 
\begin{align} \label{dipole6}
\langle J' M_{J'} | \hat{\vec{D}} | J M_J \rangle 
=& 
\delta_{SS'}(-1)^{l_2+l_1+S+J'}  \sqrt{[L][l'_2][J'][L']} \nonumber
\\ 
&\times C_{JM_{J},1q}^{J' M_{J'}}
\begin{Bmatrix}
L & S & J 
\\
J' & 1 & L'
\end{Bmatrix}
\begin{Bmatrix}
l_2 & l_1 & L 
\\
L' & 1 & l'_2 
\end{Bmatrix}                \nonumber
\\
&\times \langle n'_2l'_2  || \hat{d}_2 || n_2l_2 \rangle,
\end{align}
where $[k]=2k+1$, and the remaining reduced single-electron dipole matrix element is given by
\begin{multline} \label{reduced_mael}
\langle n'l'||\hat{d}|| nl \rangle=(-1)^{l'} \sqrt{2l+1} 
\begin{pmatrix}
l' & 1 & l
\\
0 & 0 & 0
\end{pmatrix}
\\
\times \int_0^{\infty} R_{n'l'}(r) R_{nl}(r) r^3 dr,
\end{multline}
the remaining integral includes radial parts of wave functions on the active electron.

Hence, evaluation of the radiative lifetimes  \eqref{lifetime} reduces to computation of the transition frequencies $\omega_{ij}$ and the radial integrals from \eqref{reduced_mael}. A rigorous $ab\,\,initio$ calculation of these radial integrals requires dedicated many-body methods such as multi configuration Hartree-Fock \cite{froese2022computational} or the random phase approximation with exchange \cite{amusia2013atomic}.  Instead, we use the method proposed in Ref.~\cite{kostelecky1985analytical}, based on quantum defect (QD) theory. Within this approach, the radial components of electronic wave functions are taken as hydrogenlike ones:  
\begin{multline} \label{analytic_wf}
R_{nl}(r) = \frac{1}{n^*\,^2} \sqrt{ (2Z)^3  \frac{\Gamma(n-l-I(l))}{2 \Gamma(n^*+l^*+1)}} 
\left( \frac{2Zr}{n^*} \right)^{l^*} 
\\
\times e^{-\frac{Zr}{n^*}} L_{n-l-1-I(l)}^{2l^*+1}(2Zr/n^*),
\end{multline}
where $n^*=n-\delta_l(n)$ and $l^*=l-\delta_l(n)+I(l)$ are the effective principal and angular momentum quantum numbers, respectively, $\Gamma(x)$ denotes the gamma function, and $L_n^k(x)$ are generalized Laguerre polynomials. $Z$ is an effective charge of the core, taken here $Z=1$ in all subsequent calculations.
An integer parameter $I(l)$ has an arbitrariness of choice with the following constraint: $l^*+\frac{1}{2}>0$ and $I(l) \leq n_{\text{min}} -l-1$, where $n_{\text{min}}$ is the
principal quantum number of the lowest unoccupied state with the given $l$. 
For each angular momentum $l$, the parameter $I(l)$ is chosen to best reproduce reference data.

To apply this method, we need the QD values. For highly excited Rydberg states ($n \gg 1$), the QD is approximately constant for each angular momentum $l$. For lower states, its variation can be described by the extended Rydberg-Ritz formula
    \begin{equation} \label{Extended_Rydberg-Ritz}
    \delta_l(n) = \delta_l^{(0)} + \frac{\delta_l^{(2)}}{\big(n-\delta_l^{(0)} \big)^2} + \frac{\delta_l^{(4)}}{\big(n-\delta_l^{(0)} \big)^4} +...,   
    \end{equation}
where the parameters $\{ \delta_l^{(m)} \}$ are determined by fitting to known data. For low-lying excited states, QD values are extracted directly from the Rydberg-Ritz formula
\begin{equation} \label{En_Ryd}
E_{nl} = I[^{88}\text{Sr}] - \frac{\tilde{Ry}[^{88} \text{Sr}] }{\big( n-\delta_l[n] \big)^2},
\end{equation}
where the ionization energy $I[^{88} \text{Sr}] = 1\,377\,012.72$  GHz \cite{couturier2019measurement}, the reduced Rydberg constant accounts for the finite nuclei mass $\tilde{Ry}[ ^{88}\text{Sr}]=3\,289\,821.43$ GHz. 

At room temperature $T=300$~K, we have $\tau_R^{-1} \sim 10^{14}$ sec$^{-1}$. In this regime, the BBR terms in Eq.~\eqref{lifetime} are relevant for closely spaced Rydberg states and must be accounted for during evaluation of the radiative lifetimes for large $n \gg 1$. An alternative way is to omit these BBR terms, assuming the surrounding temperature of a few Kelvin, so that $\tau_R \omega_{ij} < 1$, and therefore $n(\omega_{ij},T) \ll 1$. This is realistic in a real experimental environment supported by cryogenic apparatus~\cite{Schymik2021, Pichard2024}.
\begin{table}[h]
    \centering
    \caption{\label{RydbergRitzParameters}
    Parameters of the Rydberg-Ritz formula \eqref{Extended_Rydberg-Ritz}
    for the states of interest.}
    \begin{tabular*}{\columnwidth}{@{\extracolsep{\fill}} l l l l @{}}
        \toprule
        \toprule
        & $5sns(^3S_1)$
        & $5snp(^3P_0)$
        & $5snd(^3D_1)$ \\
        & ($15 \leq n$) \cite{couturier2019measurement}
        & ($15 \leq n$)\footnote{These values are obtained by the authors from the transition energies reported in Ref.~\cite{armstrong1979bound}}
        & ($28 \leq n$) \cite{couturier2019measurement} \\
        \midrule
        $\delta_l^{(0)}$ & $3.370\ 778$ & $2.883\ 326$ & $2.675\ 17$ \\
        $\delta_l^{(2)}$ & $0.418$ & 0.255 & $-13.15$ \\
        $\delta_l^{(4)}$ & $-0.3$ & 4.07 & $-4444$ \\
        \bottomrule
        \bottomrule
    \end{tabular*}
\end{table}

\subsection{Lifetimes of triplet states \texorpdfstring{$5sns \left( ^3S_{1} \right)$}{5sns3S1}}
The triplet states $5sns \left( ^3S_{1} \right)$ are dipole coupled to $5sn'p \left( ^3P_{J=0,1,2} \right)$ manifold. Therefore, to construct radial wave functions for these states and calculate the radiative lifetimes, we must know the QD of the corresponding Rydberg levels.

For the  $5sns \left( ^3S_{1} \right)$ series, the QDs for $n \geq 15 $ are well described by the extended Rydberg-Ritz formula \eqref{Extended_Rydberg-Ritz}, using the parameters listed in Table \ref{RydbergRitzParameters}. These parameters were derived from high-precision transition-frequency measurements reported in Ref. \cite{couturier2019measurement} (the accuracy is up to 1 MHz). For the lower-lying states $(6 \leq n \leq 14)$, the QD values are obtained directly from the Rydberg formula \eqref{En_Ryd} using the available experimental energies.

Experimental data for the $5snp \left( ^3P_{J} \right)$ states are less precise \cite{armstrong1979bound}. For $n>15$, the energies of states $5snp \left( ^3P_{0} \right)$ are not available, and the fine structure is unresolved. Thus, we proceed under the assumption that all fine-structure share the same QD. For  $n \geq 15$, the extracted QD values follow the extended Rydberg–Ritz formula \eqref{Extended_Rydberg-Ritz} with the parameters given in Table \ref{RydbergRitzParameters}. For lower-lying states $5 \leq n\leq 15$, QDs are again determined manually from available experimental energies via Eq.~\eqref{En_Ryd}.

We use these QD values to construct radial wave functions \eqref{analytic_wf}, using $I(0)=4$, $I(1)=2$ for triplet $s$ and $p$ states, respectively. The wave functions are used to calculate the radial integrals for transition dipole moments \eqref{reduced_mael}. The latter ones are used for calculations for the radiative decay rates \eqref{lifetime}. The results for the calculated radiative lifetimes are listed in Table \ref{Tab:lifetimes} for low temperatures $n(\omega_{ij},T) \to 0$, which are used in the simulations reported in the main text. 

\begin{figure}[t]
    \centering
    \includegraphics[width = 0.9\columnwidth]{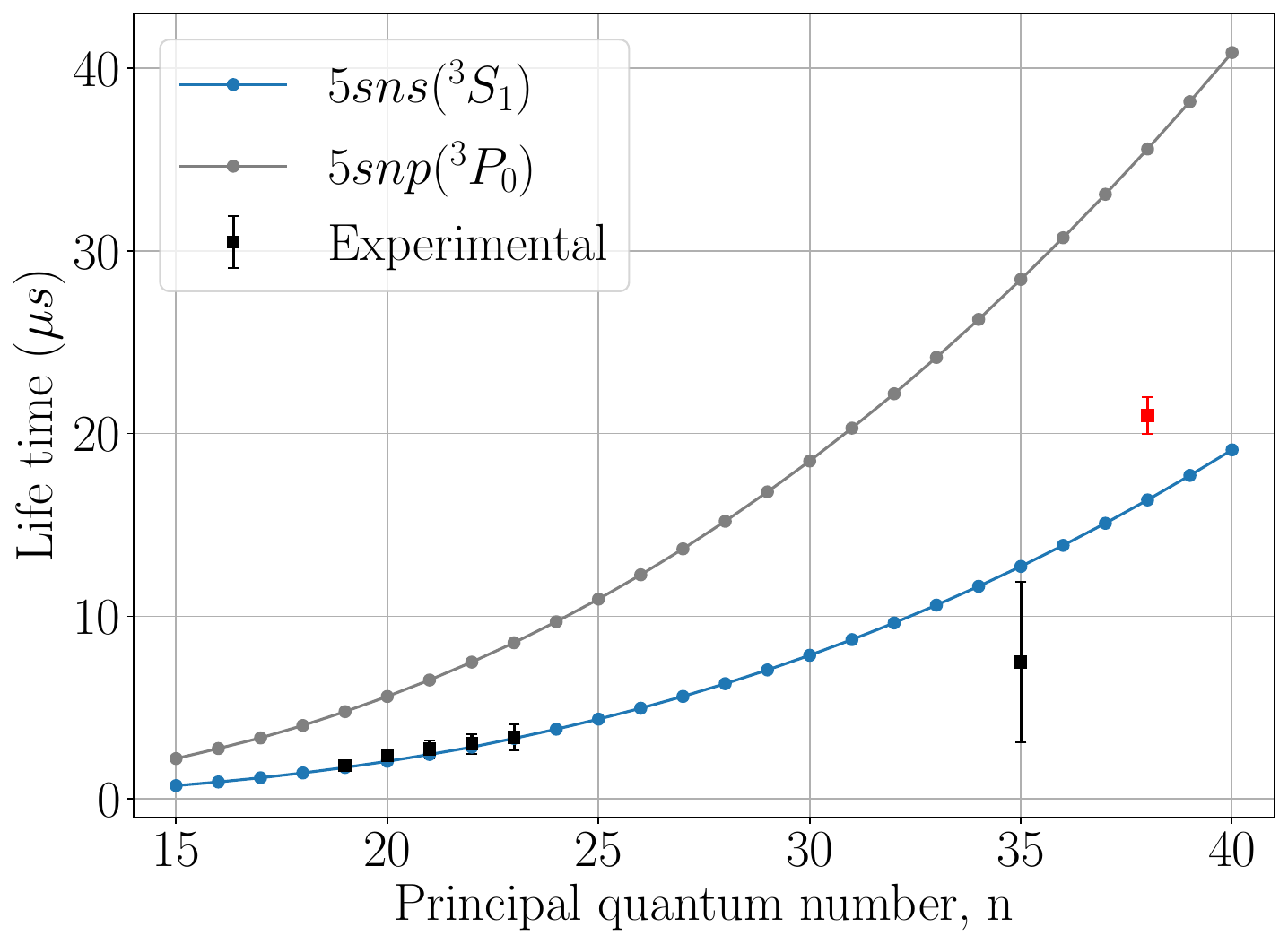}
    \caption{Radiative lifetimes $\tau_{\rm decay} =\Gamma^{-1}$ at room temperature ($T=300$~K) of triplet $5sns(^3S_1)$ and $5snp(^3P_0)$ states obtained via calculation of Eq.~\eqref{lifetime} with dipole moments evaluated with QD-based wave functions \eqref{analytic_wf}. The calculated values are shown by blue and gray dots, respectively. Black squares are the experimental results of triplet $5sns(^3S_1)$ state from Ref.~\cite{kunze1993lifetime}; the red square value is reported in Ref.~\cite{camargo2016lifetimes}. Vertical bars denote measurement uncertainties.}
    \label{fig:lifetimes}
\end{figure}
There are only limited experimental data available for the radiative lifetimes
of triplet $s$ states reported in  Refs. \cite{kunze1993lifetime, camargo2016lifetimes}. In both works, the measurements were performed at room temperature. We compare these data with our calculations provided for $T=300$~K in Fig.~\ref{fig:lifetimes}. Experimental measurements reported in Ref. \cite{kunze1993lifetime} for $19 \leq n \leq 23$ are consistent with our results within the stated uncertainties. The experimental point for $n=35$ reported in Ref.~\cite{kunze1993lifetime} and the point for $n=38$ from Ref.~\cite{camargo2016lifetimes} are noticeably inconsistent with each other. 
Our calculated values lie between the two measurements. The calculated decay rates for triplet $s$ states scale with the principal quantum number as $\Gamma_{ns(^3S_1)} \sim n^{-3.0}$ for $T=300$~K and $\Gamma_{ns(^3S_1)} \sim n^{-3.3}$ for $T=0$~K.

\subsection{Lifetimes of triplet states \texorpdfstring{$5snp \left( ^3P_{0} \right)$}{5snp3P0}}

These states have two decay channels: $|5snp \left( ^3P_{0} \right) \rangle \to |5sns \left( ^3S_{1} \right) \rangle$ and $|5snp \left( ^3P_{0} \right) \rangle \to |5snd \left( ^3D_{1} \right) \rangle$. 
The QDs of the $|5snp \left( ^3P_{0} \right)$ and $|5sns \left( ^3S_{1} \right) \rangle$ states are discussed above.

The energies of the triplet states $5snd \left( ^3D_{1} \right)$ were measured and reported in Ref. \cite{couturier2019measurement} for  $12 \leq n \leq 50$ with high precision (up to 1 MHz). For $n\geq 28$, these energies are well reproduced with the use of QDs that obey the extended Rydberg-Ritz formula \eqref{Extended_Rydberg-Ritz},  with the parameters listed in Table \ref{RydbergRitzParameters}.

\begin{table}[h]
    \centering
    \caption{\label{Energies_3D1low}
    Quantum defects for selected low-excited states
    $|5snd(^3D_{1})\rangle$. Values marked by ``*'' are obtained by
    numeric interpolation of available data.}
    \begin{tabular*}{\columnwidth}{@{\extracolsep{\fill}} l l l l @{}}
        \toprule
        \toprule
        $n$ & QD & $n$ & QD \\
        \midrule
        4  & $2.012\ 24$   & 12 & $1.911\ 64$ \\
        5  & $1.830\ 73$   & 13 & $1.979\ 45$ \\
        6  & $1.808\ 57$   & 14 & $2.082\ 10$ \\
        7  & $1.806\ 10$   & 15 & $2.214\ 70$ \\
        8  & $1.807\ 08$*  & 16 & $2.373\ 02$ \\
        9  & $1.814\ 82$*  & 17 & $2.440\ 31$ \\
        10 & $1.832\ 63$*  & 18 & $2.503\ 52$ \\
        11 & $1.863\ 80$*  & 19 & $2.544\ 44$ \\
        \bottomrule
        \bottomrule
    \end{tabular*}
\end{table}

For the lower-lying states, the QDs are extracted manually from the Rydberg formula \eqref{En_Ryd} using available experimental data: Ref. \cite{couturier2019measurement} for $n \geq 12$ and  Ref.
\cite{rubbmark1978rydberg} for $4 \leq n \leq 7$. No experimental energy data are found for $8 \leq n \leq 11$; the QDs for these states were therefore obtained here via numerical interpolation of known data of neighbor energy levels; the results are listed in Table \ref{Energies_3D1low}. These QDs were used to construct the radial wave functions \eqref{analytic_wf} of triplet $d$ states. The integer parameter entering the effective orbital quantum number is taken as $I(2)=2$, except for the first two states with $n=4$ and $n=5$, for which we take $I(2)=0$. 

The calculated radiative lifetimes for $T=0$~K are listed in Table \ref{Tab:lifetimes}; these values are also used in the analysis in Sec.~\ref{sec:noise_budget} and in Fig.~\ref{fig:iswap_Sr_Cs_noise_analysis}. Some of the values for the room temperature $T=300$~K are
shown in Fig.~\ref{fig:lifetimes}. Unfortunately, no experimental data could be found for comparison. The calculated decay rates for triplet $p$ states scale with principal quantum number  as $\Gamma_{np(^3P_0)} \sim n^{-2.7}$ for $T=300$~K and $\Gamma_{np(^3P_0)} \sim n^{-3.2}$ for $T=0$~K.  
\begin{table}[h]
    \centering
    \caption{\label{Tab:lifetimes}
    Calculated radiative lifetimes $\tau_{\rm decay}=\Gamma^{-1}$ in the
    limit $n(\omega_{ij},T) \to 0$ for some Rydberg states; the values are
    given in microseconds units.}
    \begin{tabular*}{\columnwidth}{@{\extracolsep{\fill}} l l l l l l @{}}
        \toprule
        \toprule
        $n$ & $5sns(^3S_1)$ & $5snp(^3P_0)$
        & $n$ & $5sns(^3S_1)$ & $5snp(^3P_0)$ \\
        \midrule
        15  & 0.77   & 2.52    & 60  & 90.99  & 250.84  \\
        20  & 2.28   & 7.05    & 65  & 117.23 & 322.41  \\
        25  & 5.04   & 14.93   & 70  & 148.15 & 406.46  \\
        30  & 9.42   & 27.26   & 75  & 184.11 & 504.14  \\
        35  & 15.81  & 45.04   & 80  & 225.43 & 616.18  \\
        40  & 24.58  & 69.27   & 85  & 272.51 & 743.72  \\
        45  & 36.10  & 100.96  & 90  & 325.72 & 887.75  \\
        50  & 50.76  & 141.12  & 95  & 385.43 & 1049.29  \\
        55  & 68.92  & 190.74  & 100 & 452.03 & 1229.33  \\
        \bottomrule
        \bottomrule
    \end{tabular*}
\end{table}

\subsection{Lifetimes of metastable states \texorpdfstring{$5s5p \left( ^3P_{0} \right)$}{5s5p3P0} and \texorpdfstring{$5s5p \left( ^3P_{2} \right)$}{5s5p3P2}}

The state $5s5p \left( ^3P_{0} \right)$ 
can potentially decay only into the physical ground state $5s^2(^1S_0)$. The value of the corresponding radiative decay rate  has been reported in Ref. \cite{muniz2021cavity}, where the measurements were performed for fermionic isotope $^{87}$Sr:
\begin{equation}
\frac{1}{2\pi} \Gamma[5s5p (^3P_0) ] = \, 1.35 \times 10^{-6}~\text{kHz}.   \nonumber
\end{equation}

The state $5s5p \left( ^3P_{2} \right)$ has the highest energy among all states of the spin-orbit manifold $5s5p \left( ^3P_{J} \right)$. Thus, it can undergo radiative decay into the lower spin-orbit states with $J=1,0$ as well as into the physical ground state $5s^2(^1S_0)$. The corresponding radiative lifetime resulting from all these processes has been reported in theoretical work \cite{derevianko2001feasibility}
\begin{equation}
\frac{1}{2\pi}\Gamma[5s5p (^3P_2) ] =  1.5 \times 10^{-7}~\text{kHz}.   \nonumber
\end{equation}
The radiative lifetimes of these metastable states are $5-7$ orders of magnitude longer than those of the Rydberg states. For this reason, we do not account for the decays of these metastable states in the noise modeling \eqref{Decay_ham}.

\section{Noise model}
\label{sec:A/noise_model}

\subsection{Interaction noise} 

Atomic motion due to finite temperature leads to position fluctuations in optically trapped atoms, which in turn induce fluctuations in the DDI between two atoms. Since the DDI strength determines the exchange frequency, any deviation from the target value $V_{\rm dipole}$ results in non-optimal Rydberg population and consequently reduces the overall gate fidelity.

The positions of two atoms, $(x_1, y_1, z_1)$ and $(x_2, y_2, z_2)$, fluctuate as $\tilde{x}_i = x_i + \delta x_i$, $\tilde{y}_i = y_i + \delta y_i$, and $\tilde{z}_i = z_i + \delta z_i$, where $i = 1, 2$. If we approximate the optical tweezer potential as a harmonic oscillator, each displacement $\delta_\alpha$ for $\alpha = \{x, y, z\}$ follows a Gaussian distribution with the standard deviation
\begin{equation}
\sigma_{\mathrm{pos}}^{\alpha}
= \sqrt{\frac{\hbar}{2 m \omega_{\alpha}}(1 + 2 \bar{n}_{\alpha})},
\label{eq:position_std}
\end{equation}
where
\begin{equation}
\bar{n}_{\alpha} =
\frac{1}{e^{\hbar \omega_{\alpha}/k_B T} - 1}.
\end{equation}
Here, $k_B$ is the Boltzmann constant,  $T$ is the temperature, $m$ is the atomic mass, and $\omega_{\alpha}$ is the trap frequency. In typical tweezer geometries, we assume $\omega_x = \omega_y \equiv \omega_{xy}$.

The noisy interatomic separation is then
\begin{equation}
\tilde{R}
= \sqrt{
(\tilde{x}_2 - \tilde{x}_1)^2
+ (\tilde{y}_2 - \tilde{y}_1)^2
+ (\tilde{z}_2 - \tilde{z}_1)^2 },
\end{equation}
which induces fluctuations in the dipole-dipole and vdW interaction strengths,
\begin{align}
    H_{\rm exchange} = &\frac{C_3}{\tilde{R}^3} \left(\ket{r r'}\bra{r' r} + \ket{r' r}\bra{r r'}\right),
     \label{eq:dipole_hamiltonian_int_noise}
\end{align}

\begin{align}
    H_{\rm vdW} = &\frac{C^{rr}_6}{\tilde{R}^6} \ket{r r}\bra{r r} + \frac{C^{r'r'}_6}{\tilde{R}^6} \ket{r' r'}\bra{r' r'} \nonumber \\
    + &\frac{C^{rr'}_6}{\tilde{R}^6}  \left( \ket{r r'}\bra{r r'} + \ket{r' r}\bra{r' r}\right) \, ,
     \label{eq:vdw_hamiltonian_int_noise}
\end{align}

As a representative example, we choose radial and axial trap frequencies of $\omega_{xy} = 2\pi \times 200$~kHz and $\omega_z = 2\pi \times 40$~kHz, which are realistic in the current experimental system. For $^{133}$Cs atoms, at a temperature of $T = 1\ \si{\micro\kelvin}$, one obtains a position standard deviation of  $\sigma_{\rm{pos}}^{\alpha}= (0.013,\ 0.013,\ 0.04)\ \si{\micro\meter}$ for each spatial dimension.

\subsection{Doppler noise}

Atomic motion not only perturbs the DDI, but also generates unwanted detuning stemming from the Doppler shift, which is known as Doppler noise. A finite velocity 
\begin{equation}
    \Delta v_{\alpha} = \sqrt{\frac{\hbar \omega_{\alpha}}{2 m}(1 + 2 \bar{n}_{\alpha})},
    \label{eq:velocity_std}
\end{equation}
shifts the resonance for each atomic level $a = \{1, r, r^\prime\}$. This shift is characterized by a Gaussian standard deviation in detuning as follows:
\begin{align}
    \sigma_{\mathrm{det}}^{a} &= k_{\mathrm{eff}}^{\alpha, a}\ \Delta v_{\alpha} \\
    &= k_{\mathrm{eff}}^{x, a}\ \Delta v_{x} + k_{\mathrm{eff}}^{y, a}\ \Delta v_{y} + k_{\mathrm{eff}}^{ z, a}\ \Delta v_{z}, \nonumber
    \label{eq:detuning_std}
\end{align}
where $k_{\mathrm{eff}}^{a}$ represents the effective wave vector for each laser coupled with the atomic level $a$.
Finally, the Doppler shifts $\Delta^a$ are sampled from the Gaussian distribution, which adds the following detuning term to the Hamiltonian:
\begin{align}
    \frac{\tilde{H}_{\rm Doppler}}{\hbar} = - \sum_{i} \bigg(\Delta^{1}(t) \ket{1}_i\bra{1} &+ \Delta^{r}(t) \ket{r}_i\bra{r} \nonumber \\
    &+ \Delta^{r'}(t) \ket{r'}_i\bra{r'} \bigg).
\end{align}

For example, take the scheme A, $\ket{1} \leftrightarrow \ket{r}$ transition in $^{133}$Cs, where $\ket{1} = \ket{6sS_{1/2}, F=4}$ and $\ket{r} = \ket{61S_{1/2}}$ case, with transition wavelength $\lambda = 319~$nm and effective wave vector $k^x_{\text{eff}} = (2\pi)3.13\times 10^6~\text{m}^{-1}$ and $k^y_{\text{eff}} = k^z_{\text{eff}} = 0$. Using the values from the previous subsection, $\omega_{x} = \omega_{y} = 2\pi \times 200$ kHz, $\omega_{z} = 2\pi \times 40$ kHz, and $T = 1\ \si{\micro\kelvin}$, one obtains a detuning standard deviation of $\sigma_{\textrm{det}}^r = k_{\textrm{eff}}^r \Delta v = 2\pi \cross 54$ kHz.

\subsection{Decay noise}
We consider an approximate method to implement the finite lifetimes of the Rydberg states $\ket{r}$ and $\ket{r'}$ into our modeling. 
We considered the non-Hermitian Hamiltonian~\cite{Pagano2022} 
\begin{align}
    \frac{\tilde{H}_{\rm decay}}{\hbar} = - \frac{i}{2} \sum_{i} \bigg(\Gamma^{\text{eff}}_{r} \ket{r}_i\bra{r}_i +  \Gamma^{\text{eff}}_{r^\prime} \ket{r'}_i\bra{r'}_i \bigg) \, ,
     \label{eq:dec_hamiltonian}
\end{align}
where $\Gamma_{r,r'}^{\text{eff}}$ represent the effective decay rates of the Rydberg states (see also Appendix~\ref{sec:App/FSQubitDecay}).
This approximation circumvents the need to solve the full Lindblad master equation with Rydberg decay jump operators, while maintaining sufficient accuracy.

\subsection{Laser phase and intensity noise} \label{App:phase}
The laser drive couples the two states with a strength coefficient $\Omega e^{i\phi}$, where $\phi$ is the phase and $\Omega$ is the Rabi frequency. Due to the finite stability of the laser, both the phase and the intensity fluctuate as $\phi \rightarrow \tilde{\phi}$ and $\Omega \rightarrow \tilde{\Omega}$~\cite{deLeseleuc2018}. We model these fluctuations as~\cite{Jiang2022}:
\begin{align}
    \tilde{\phi} (t) &= \phi(t) + \delta \phi(t), \label{eq:noisy_phases}\\
    \tilde{\Omega} (t) &= \Omega(t) \left( 1 + \frac{\alpha_I(t)}{2} \right), \label{eq:noisy_intensities}
\end{align}
where $\delta\phi$ and $\alpha_I$ denote the phase noise and the relative intensity noise. These noise contributions are quantified by their respective PSDs, $S_\phi$ and $S_{\alpha_I}$. Accordingly, we express them as
\begin{align}
    \delta \phi(t) &= \sum_{f} 2 \sqrt{S_{\phi}(f) \Delta f} \cos{(2 \pi f t + \varphi_f)} \label{eq:phase_PSD}, \\
    \alpha_I(t) &= \sum_{f} 2 \sqrt{S_{\alpha_I}(f) \Delta f} \cos{(2 \pi f t + \varphi_f)}, \label{eq:intensity_PSD}
\end{align}
where $\Delta f = f_{j+1} - f_j$ is the spacing between sampled frequency components, and each $\varphi_f$ is a random phase drawn from a uniform distribution $[0, 2\pi]$ for each frequency component $f = f_j$~\cite{Jiang2022}.

These fluctuations are incorporated into the dynamics through the driving Hamiltonian,
\begin{align}
    \frac{\tilde{H}_{\rm drive}^{(A)}(t)}{\hbar} = \sum_{i} \bigg[ \bigg( &\frac{1}{2} \tilde{\Omega}^{1r}(t) \E^{\I \tilde{\phi}^{1r}(t)} \ket{r}_i\bra{1} \nonumber\\
     + &\frac{1}{2} \tilde{\Omega}^{0r'}(t) \E^{\I \tilde{\phi}^{0r'}(t)} \ket{r'}_i\bra{0}  + \mathrm{H.c.} \bigg),
\end{align}
\begin{align}
    \frac{\tilde{H}_{\rm drive}^{(B)}(t)}{\hbar} = \sum_{i} \bigg[ \bigg( &\frac{1}{2} \tilde{\Omega}^{01}(t) \E^{\I \tilde{\phi}^{01}(t)} \ket{1}_i \bra{0}  \nonumber\\
     + &\frac{1}{2} \tilde{\Omega}^{1r}(t) \E^{\I \tilde{\phi}^{1r}(t)} \ket{r}_i\bra{1} \nonumber\\
     + &\frac{1}{2} \tilde{\Omega}^{rr'}(t) \E^{\I \tilde{\phi}^{rr'}(t)} \ket{r'}_i\bra{r}  + \mathrm{H.c.} \bigg).
\end{align}

The impact of phase noise can be further reduced, at the cost of decreasing laser power, through cavity filtering for a given full width at half maximum power linewidth of the cavity, $f_c$. The filtered PSD,
\begin{equation}
    S_{\text{filt}}(f) = |H(f)|^2 \times S(f),
    \label{eq:cavity_filtering}
\end{equation}
depends on the cavity amplitude transfer function,
\begin{equation}
    H(f) = \frac{1}{1 + 2\I(f - f_0)/\Delta f_{\rm{c}}}.
\end{equation}
We assume the linewidth $f_{\rm c}$ to be $140$ kHz, as reported in Ref.~\cite{Endres2025Benchmark}.

\section{Fidelity Response Theory on Laser noise} \label{App:FRT}

FRT~\cite{Endres2025Benchmark} is a powerful framework for analyzing and predicting the sensitivity of gate fidelity to noise.
We apply FRT to frequency and intensity noise. We consider these two noises as a perturbation,  
\begin{equation}
    \hat{H}(t) = \hat{H}_0(t) + \sum_{j} h_j(t) \hat{O}_j(t),
\end{equation}
where $\hat{H}_0(t)$ is the noiseless Hamiltonian, $\hat{O}_j(t)$ is a noise operator that encodes information about the noise type, and $h_j(t)$ is a function that encodes the amplitude of the noise. $h_j(t)$ depends on the PSD $S_j(f)$, where $h(t) = \sum_{f} 2 \sqrt{S (f) \Delta f} \cos{(2 \pi f t + \varphi_f)} $. The fidelity is derived from FRT by using the first-order perturbation,
\begin{equation}
    1 - \mathcal{F} = \int_0^{\infty} df S(f) I(f, \Omega),
\end{equation}
where $I(f)$ denotes the RF, 
\begin{align}
    I(f) = \int_0^T \int_0^T dt d\tau &\cos{(2\pi f (t-\tau))} \nonumber \\
    &\times \langle \hat{O}^H(t, \Omega) \hat{O}^H(\tau, \Omega)\rangle_c.
    \label{eq:FRT_resp_fun}
\end{align}
$\langle \hat{A} \hat{B} \rangle_c = \langle \hat{A} \hat{B} \rangle - \langle \hat{A} \rangle \langle \hat{B} \rangle$ represents the connected correlator function, where $\langle \hat{A} \rangle$ is short for $\bra{\Psi_0} \hat{A} \ket{\Psi_0}$. $\hat{O}^H(t)$ represents the noise operator $\hat{O}(t)$ in the Heisenberg picture $\hat{O}^H(t) = \hat{U}^{\dagger}(t) \hat{O}(t) \hat{U}(t)$ for the unitary evolution operator $\hat{U}(t)$ under $\hat{H}_0(t)$.

\subsection{Frequency noise}
For fluctuations in the frequency of the lasers, we move to the rotating wave (RW) frame and derive the following time-independent noise operators:
\begin{equation}
    \hat{O}_{\nu^{a}} = -2\pi \sum_i \ket{a}_i\bra{a}.
\end{equation}
The infidelity then becomes
\begin{equation}
    1 - \mathcal{F}_{\nu} = \int df S_{\nu}(f) I_{\nu}(f) \ ,
\end{equation}
with the frequency response function
\begin{align}
    I_{\nu}(f) = \sum_{a} \int_0^T \int_0^T &dt\, d\tau\, \cos\big(2\pi f (t - \tau)\big)\, \nonumber \\
    &\cross \left\langle \hat{O}_{\nu^a}^H(t)\, \hat{O}_{\nu^a}^H(\tau) \right\rangle_c \ .
\end{align}

\subsection{Intensity noise}
For intensity noise, the derivation is slightly simpler since the driving term with intensity/Rabi frequency noise can be separated into noisy and noiseless terms [see \eqref{eq:noisy_intensities}] without any approximate expansions,
\begin{align}
    \tilde{H}_{\mathrm{drive}}(t) &= \sum_{ab} \sum_i \frac{\tilde{\Omega}^{ab}(t)}{2} \left( \E^{\I \phi^{ab}(t)} \ket{b}_i \bra{a} + \mathrm{H.c.} \right) \nonumber \\
    &= \hat{H}_{\mathrm{drive}}(t) + \sum_{ab} \alpha^{ab}_I (t) \hat{O}_{\Omega^{ab}}(t),
\end{align}
where the relative intensity noise $\alpha^{ab}_I (t)$ is given by Eq.~\eqref{eq:intensity_PSD}, leading to the following noise operators:
\begin{equation}
    \hat{O}_{\Omega^{ab}}(t) = \frac{\Omega^{ab}(t)}{4} \sum_i \left( \E^{\I \phi^{ab}(t)} \ket{b}_i \bra{a} + \mathrm{H.c.} \right).
\end{equation}
Therefore, the infidelity becomes
\begin{equation}
    1 - \mathcal{F}_{\Omega} = \int df S_{\alpha_I}(f) I_{\Omega}(f, \Omega) \ ,
\end{equation}
with the intensity noise response function
\begin{align}
    I_{\Omega}(f)= \sum_{ab} \int_0^T \int_0^T & dt\, d\tau\, \cos\big(2\pi f (t - \tau)\big)\, \nonumber \\
    & \cross \left\langle \hat{O}_{\Omega^{ab}}^H(t)\, \hat{O}_{\Omega^{ab}}^H(\tau) \right\rangle_c \ .
\end{align}

\end{document}